\documentclass[twocolumn,11pt,twoside]{article}
\usepackage{epsfig}

\def\check#1{}
\def\modell{{\sl poly--gonato} model}
\def\lna{\langle\ln A\rangle}

\begin{document} 
\twocolumn[{\centering
 \vspace*{3cm}
{\LARGE\bf On the Knee in the Energy Spectrum of Cosmic Rays\\[5mm]}

{\Large J\"org R. H\"orandel\\[3mm]}
{\small\sl University of Karlsruhe, Institut f\"ur Experimentelle Kernphysik,
            P.O. Box 3640, 76\,021 Karlsruhe, Germany\\
	    http://www-ik.fzk.de/$\sim$joerg\\[5mm]}

Preprint arXiv:astro--ph/0210453\\ 
Submitted to Astroparticle Physics 15. April 2002; accepted 6. August
2002.\\[5mm]
}

\hspace*{0.1\textwidth}\begin{minipage}{0.8\textwidth}
\hrulefill\\
{\bf Abstract\\}
{\small
 The {\sl knee} in the all--particle energy spectrum is scrutinized with a
 phenomenological model, named \modell, linking results from direct and
 indirect measurements.  For this purpose, recent results from direct and
 indirect measurements of cosmic rays in the energy range from 10~GeV up to
 1~EeV are examined.  The energy spectra of individual elements, as obtained by
 direct observations, are extrapolated to high energies using power laws and
 compared to all--particle spectra from air shower measurements.  A cut--off
 for each element proportional to its charge $Z$ is assumed.  The model
 describes the {\sl knee} in the all--particle energy spectrum as a result of
 subsequent cut--offs for individual elements, starting with the proton
 component at 4.5~PeV, and the second change of the spectral index around
 0.4~EeV as due to the end of stable elements $(Z=92)$.  The mass composition,
 extrapolated from direct measurements to high energies, using the \modell, is
 compatible with results from air shower experiments measuring the
 electromagnetic, muonic, and hadronic components. But it disagrees with the
 mass composition derived from $X_{max}$ measurements using \v{C}erenkov and
 fluorescence light detectors. \\
}
{\small\sl Key words:} Cosmic rays; Energy spectrum; Knee; Mass composition; Air shower\\  
{\small\sl PACS:} 96.40.De, 96.40.Pq, 98.70.Sa\\
\hrule
\end{minipage} 

\vspace*{5mm}
]

\section{Introduction}
After almost one century of cosmic--ray physics, a statement made by V. Hess in
his Nobel Lecture in 1936 is still true: {\sl From a consideration of the
immense volume of newly discovered facts in the field of physics $\ldots$ it
may well appear $\ldots$ that the main problems were already solved $\ldots$.
This is far from being truth, as will be shown by one of the biggest and most
important newly opened fields of research, $\ldots$ that of cosmic rays}
\cite{hessnobel}.  Starting with very simple instruments flown in manned
balloon gondolas in 1912 \cite{hess} the field developed to present--day
experiments, using  modern particle physics detectors at ground level, in the
upper atmosphere, and in space to detect cosmic--ray particles.  Many
properties of cosmic rays have been clarified by these measurements, but the
origin of the relativistic particles is still under discussion.  Since charged
particles are deflected by the irregular galactic magnetic fields, the arrival
direction cannot be used to identify their sources, but measurements of their
energy spectra allow to draw conclusions about their origin.

The energy spectra for individual elements are obtained directly with satellite
and balloon--borne experiments at the top of the atmosphere up to several
$10^{13}$~eV and for groups of elements up to $10^{15}$~eV.  Due to the fast
decreasing flux, measurements at higher energies require large detection areas
or long exposure times, which presently can only be realized in ground--based
detector systems. These experiments measure extensive air showers, generated by
interactions of the high energetic cosmic rays with the nuclei in the
atmosphere.

\begin{figure}[hbt]
 \epsfig{file=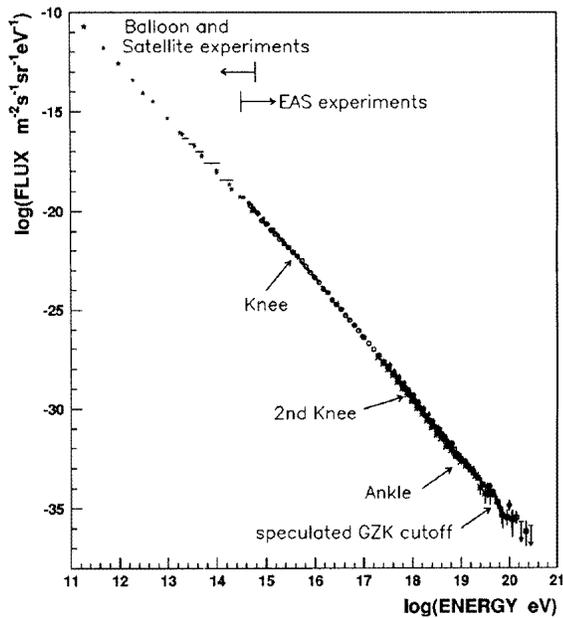,width=\columnwidth}
 \caption{\small Cosmic--ray energy spectrum, taken from \cite{watson}.}
 \label{watson}
\end{figure}

The primary energy spectrum extends over many orders of magnitude from GeV
energies up to at least $10^{20}$~eV.  It is a steeply falling power law
spectrum with almost no structure as can be inferred from Figure~\ref{watson}.
The only prominent features are a change in the spectral index at about 4~PeV,
generally called {\sl the knee}, a slight steepening around 400~PeV, designated
as 2nd {\sl knee} in Figure~\ref{watson}, and a flattening at the highest
energies around 10~EeV, called {\sl the ankle}.  The {\sl knee} was first
observed by the MSU group in the electromagnetic component more than 40~years
ago \cite{kulikov}.  Since then it has been confirmed by many groups in other
components of extensive air showers, viz. the electromagnetic, muonic, and
hadronic component, as well as in their \v{C}erenkov radiation.  Yet its origin
is still under discussion and is generally believed to be a corner stone in
understanding the origin of cosmic rays.

In this article the two--{\sl knee} structure is explained as a consequence of
the all--particle spectrum being composed of individual spectra of elements
with distinguished cut--offs. The first and prominent {\sl knee} is due to the
subsequent cut-offs for all elements, starting with the proton component, and
the second {\sl knee} marks the end of the stable elements ($Z=92$).

The idea of composing the all--particle spectrum in the {\sl knee} region out
of individual elemental spectra has been discussed in the literature, e.g.
Webber sketches the {\sl knee} in the energy spectrum as result of a rigidity
dependent cut--off for individual elements \cite{webberknee} or Stanev et al.
calculate the all-particle spectrum as sum spectrum for groups of elements from
hydrogen to iron \cite{stanev}.

In the following the phenomenological model mentioned is described, for which
the Greek notation {\sl poly gonato}\footnote{Greek "many knees"} is chosen.
In section~\ref{direct} and \ref{indirect} an overview is given of existing
direct measurements of individual nuclei up to ultra--heavy elements which are
important to explain the second {\sl knee}, as well as of indirect measurements
of the all--particle spectrum.

Preliminary results of this work have been presented earlier \cite{jrhicrc}.
Since then, some parameters have changed due to inclusion of recent results and
application of more detailed analysis procedures.

\section{The Phenomenological Model}
Cosmic--ray particles are most likely accelerated in strong shock fronts of
supernova remnants by diffusive shock acceleration.  The mechanism first
introduced by Fermi \cite{fermi} has been improved by many authors, see e.g.
\cite{berezhko,biermann,drury,ellisonsnr,blandford}.  In these models particles
are deflected by moving magnetized plasma clouds, the particles cross the shock
front many times, and thereby gain energy up to the PeV region and achieve a
nonthermal energy distribution.  The Fermi acceleration at strong shocks
produces a power--law spectrum close to that observed.

During the diffusive propagation of the particles through the galaxy on a
random path deflected by turbulent magnetic fields, the energy spectrum of the
particles is modified.  The differences between source spectra and observed
spectra have been explained due to nuclear spallation, leakage from the galaxy,
nuclear decay, ionization losses, and, for low energies, solar modulation, see
e.g. \cite{blandford,cesarsky,letaw}.

The \modell\ parametrizes the energy spectra of individual elements, based on
measured data, assuming power laws and taking into account the solar modulation
at low energies.

At energies in the GeV region the solar modulation of the energy spectra is
described using the parametrization
\begin{equation}\label{solmod}
 \begin{array}{rl}
 \frac{\textstyle d\Phi_Z}{\textstyle dE}(E)= 
    & N \frac{\textstyle E(E+2 m_A)}{\textstyle E+M} \cdot\\
    & \frac{\textstyle \left(E+M+780\cdot e^{-2.5\cdot10^4 E}\right)^{\gamma_Z}}
           {\textstyle E+M+2 m_A}\\
 \end{array}   
\end{equation}
adopted from \cite{bonino}.  $N$ is a normalization constant, $\gamma_Z$ the
spectral index of the anticipated power law at high energies (see below),
$E=E_0/A$ the energy per nucleon, $m_A$ the mass of the nucleus with mass
number $A$, and $M$ the solar modulation parameter.  A value $M=750$~MeV is
used for the parametrization.

Above about $Z\cdot 10$ GeV, the modulation due to the magnetic field of the
heliosphere is negligible and the energy spectra of cosmic--ray nuclei are
assumed to be described by power laws.  The finite life time of supernova
blast waves limits the maximum energy $E_{max}$ to be obtained during the
acceleration process. The latter becomes inefficient at an energy $\hat{E}_Z
\propto Z$, where $Z$ is the nuclear charge of the particle, and a complete
cut--off is expected at the maximum energy $E_{max}$, also proportional to $Z$,
see e.g. \cite{wiebel,berezhkomax,lagagemax,voelkmax}.  Therefore, a cut--off
or at least a change of the slope in the spectrum for the individual species at
an energy $\hat{E}_Z\propto Z$ is expected.

In the leaky box model for cosmic--ray propagation particles are contained in a
well defined volume of the galaxy, but have a small probability to escape from
this volume.  The gyromagnetic radius for a particle with charge $Z$, energy
$E_{15}$ in PeV, and momentum $p$ in a magnetic field $B$ is given by $R=p/e Z
B \approx (10~\mbox{pc})\,E_{15}/Z B\,(\mu\mbox{G})$ \cite{wandel}.  In a
simple picture of the propagation low $Z$ particles are more likely to escape
from the galaxy as compared to particles with high $Z$ at the same energy due
to their larger gyromagnetic radii.

In diffusion models, the global toroidal magnetic field in the galaxy disturbs
the random walk of particles in the stochastic magnetic fields and causes a
systematic drift or Hall diffusion \cite{ptuskin}.  This drift is rapidly
rising with energy. It dominates at the {\sl knee} the slowly increasing usual
diffusion and leads to a strong leakage from the galaxy.  Diffusion and drift
of different nuclei at ultrarelativistic energies depend on the energy per unit
charge $E_0/Z$, and consequently lead to a rigidity dependent cut--off for
individual elements.

To summarize, both scenarios --- acceleration and propagation --- entail that
the cut--off energy for each individual element depends on its charge $Z$.
Inspired by these theories the following ansatz is adopted to describe the
energy dependence of the flux for particles with charge $Z$ 
\begin{equation}\label{specfun}
  \frac{d\Phi_Z}{dE_0}(E_0) = \Phi_Z^0 E_0^{\gamma_Z}
          \left[1+\left(\frac{E_0}{\hat{E}_Z}\right)^{\textstyle\epsilon_c}\right]
          ^\frac{\textstyle\gamma_c-\gamma_Z}{\textstyle\epsilon_c}.
\end{equation}
The absolute flux $\Phi_Z^0$ and the spectral index $\gamma_Z$ quantify the
power law. The flux above the cut--off energy is modeled by a second and
steeper power law.  $\gamma_c$ and $\epsilon_c$ characterize the change in the
spectrum at the cut--off energy $\hat{E}_Z$.  Both parameters are assumed to be
identical for all spectra, $\gamma_c$ being the hypothetical slope beyond the
{\sl knee} and $\epsilon_c$ describes the smoothness of the transition from the
first to the second power law. $\epsilon_c=1$ corresponds to a smooth change in
the range of about one decade of energy, larger values describe a faster
change, e.g.  $\epsilon_c=4$ leads to a change within 1/5 of a decade.

To study systematic effects, instead of a common spectral index for all
elements above the cut--off energy also a constant difference $\Delta\gamma$
between the spectral indices below and above the {\sl knee} is tried and the
spectrum for an element with charge $Z$ is assumed as
\begin{equation}\label{specfundg}
  \frac{d\Phi_Z}{dE_0}(E_0) = \Phi_Z^0 E_0^{\gamma_Z}
       \left[1+\left(\frac{E_0}{\hat{E}_Z}\right)^{\textstyle\epsilon_c}\right]
       ^\frac{\textstyle-\Delta\gamma}{\textstyle\epsilon_c}.
\end{equation}

From an astrophysical point of view a rigidity dependent cut--off
$\hat{E}_Z\propto Z$ is the most likely description, as motivated above.  But,
since the origin of the {\sl knee} is still under discussion, also other
explanations for the change of the spectral slope are possible.  Some theories
still favour a particle physical origin due to a change of the interactions in
the atmosphere (see e.g. \cite{nikolsky}).  Such effects would probably depend
on the energy per nucleon $E_0/A$ and lead to a nuclear mass dependent cut--off
$\hat{E}_Z\propto A$.  The last possibility considered in this article is, that
the bend in the individual spectra appears at a constant energy.  In the
\modell\ all three assumptions to parametrize the cut--off energy are
scrutinized.
\begin{equation}\label{cutofffun}
 \hat{E}_Z=\left\{
 \begin{array}{ll}
  \hat{E}_p\cdot Z &;\mbox{rigidity dependent}\\
  \hat{E}_p\cdot A &;\mbox{mass dependent}\\
  \hat{E}_p        &;\mbox{constant}\\
 \end{array}\right.
\end{equation} 
It will be investigated with measured data, which ansatz describes the data
best.  In the following, only "\modell" refers to the rigidity dependent
hypothesis.

Knowing the flux $d\Phi_Z/dE_0(E_0)$ for the species with charge $Z$, the flux
of the all--particle spectrum is obtained by summation over all cosmic--ray
elements
\begin{equation}\label{allpartfun}
  \frac{d\Phi}{dE_0}(E_0)=\sum_{Z=1}^{92} \frac{d\Phi_Z}{dE_0}(E_0) 
  \quad\mbox{.}
\end{equation}
This equation contains the parameters $\Phi_Z^0$ and $\gamma_Z$ for each
element and the common parameters, $\hat{E}_p$, $\epsilon_c$, and $\gamma_c$ or
$\Delta\gamma$.  $\Phi_Z^0$ and $\gamma_Z$ are obtained from direct
measurements of individual nuclei. The remaining three parameters are derived
from a fit to the all--particle spectrum as obtained by indirect measurements.

\section{Direct Measurements}\label{direct}
\subsection{Elements with $Z\le28$}
\begin{figure}[hbt]
 \epsfig{file=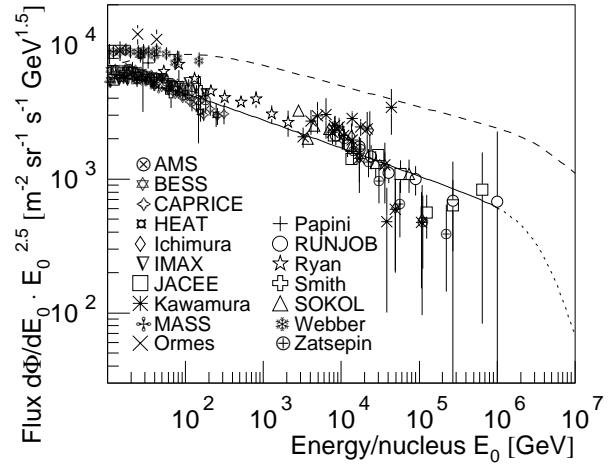,width=\columnwidth}
 \caption{\small Differential energy spectrum for protons.
	  The best fit to the spectrum according to a power law is represented
	  by the solid line, the bend (dotted line) is obtained from a fit to
	  the all--particle spectrum, see section~\ref{espec}. The
	  all--particle spectrum is shown as dashed line for reference.
	  Results are shown from
   AMS \cite{ams},
   BESS  \cite{bess},
   CAPRICE \cite{caprice},
   HEAT \cite{heat},
   Ichimura et al. \cite{ichimura},
   IMAX \cite{imax},
   JACEE \cite{jaceephe},
   Kawamura et al. \cite{kawamura},
   MASS \cite{mass},
   Ormes et al. \cite{ormes},
   Papini et al. \cite{papini},
   RUNJOB \cite{runjobphe},
   Ryan et al. \cite{ryan},
   Smith et al. \cite{smith},
   SOKOL \cite{sokol},
   Webber et al. \cite{webber}, and
   Zatsepin et al. \cite{zatsepin}.}
 \label{directP}
\end{figure}

\begin{figure}[hbt]
 \epsfig{file=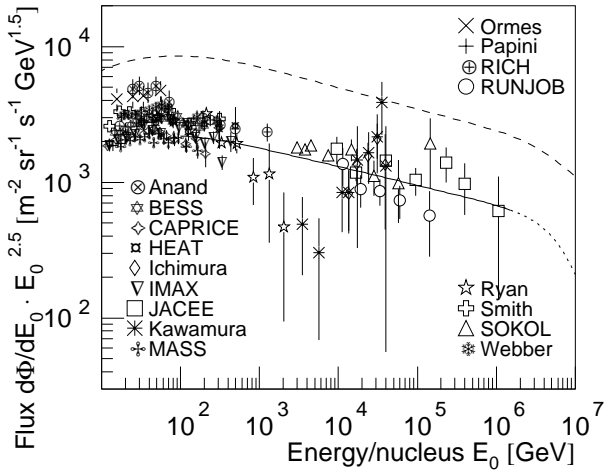,width=\columnwidth}
 \caption{\small Differential energy spectrum for helium nuclei.
	  The best fit to the spectrum according to a power law is represented
	  by the solid line, the bend (dotted line) is obtained from a fit to
	  the all--particle spectrum, see section~\ref{espec}.  The
	  all--particle spectrum is shown as dashed line for reference.
   Used are values from
   Anand et al. \cite{anand},
   BESS  \cite{bess},
   CAPRICE \cite{caprice},
   HEAT \cite{heat},
   Ichimura et al. \cite{ichimura},
   IMAX \cite{imax},
   JACEE \cite{jaceephe},
   Kawamura et al. \cite{kawamura},
   MASS \cite{mass},
   Ormes et al. \cite{ormes},
   Papini et al. \cite{papini},
   RICH \cite{rich},
   RUNJOB \cite{runjobphe},
   Ryan et al. \cite{ryan},
   Smith et al. \cite{smith},
   SOKOL \cite{sokol}, and
   Webber et al. \cite{webber}.} 
 \label{directHe}
\end{figure}

\begin{figure}[hbt]
 \epsfig{file=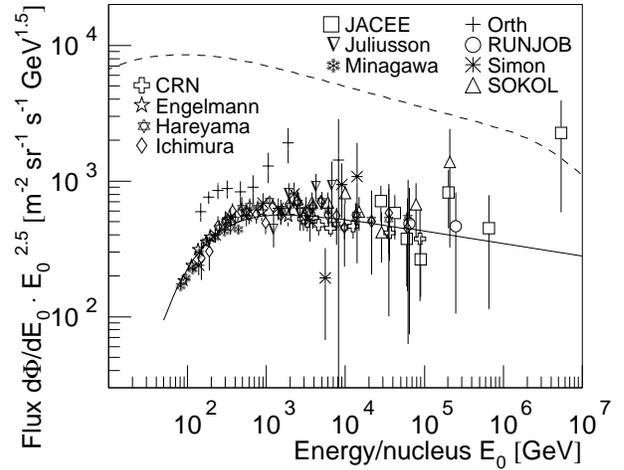,width=\columnwidth}
 \caption{\small Differential energy spectrum for iron nuclei. 
          The best fit to the spectrum is represented by a solid line,
          the all--particle spectrum is shown as dashed line for reference.
   Plotted are measurements according to	  
   CRN   \cite{crn},          
   Engelmann et al. \cite{engelmann},
   Hareyama et al. \cite{hareyama},
   Ichimura et al. \cite{ichimura},
   JACEE \cite{jaceefe},
   Juliusson et al. \cite{juliusson},
   Minagawa et al. \cite{minagawa},
   Orth et al.  \cite{orth},
   RUNJOB \cite{runjoball},
   Simon et al.  \cite{simon}, and
   SOKOL \cite{sokol}.}
 \label{directFe}
\end{figure}

Many direct measurements of the energy spectra of individual cosmic--ray nuclei
have been performed.  Figures~\ref{directP} to \ref{directFe} show compilations
of results for protons, helium and iron nuclei.

The differential flux for the most abundant cosmic--ray element --- hydrogen
--- multiplied by $E_0^{2.5}$ is plotted in Figure~\ref{directP} versus the
particle energy $E_0$.  The data show a decreasing flux as function of energy.
The straight line represents the fit according to a power law.  The errors as
specified by the different experiments have been taken into account.  Best fit
values are $\Phi_p^0=(8.73\pm0.07)\cdot10^{-2}~$~(m$^2$\,s\,sr\,TeV)$^{-1}$ and
$\gamma_p=-2.71\pm0.02$ (solid curve) with $\chi^2\mbox{/d.o.f.}=3.4$.  The
errors specify the statistical uncertainties.  The cut--off shown in the figure
as dotted line will be discussed in section \ref{espec} when fitting the
all--particle spectrum within the \modell.  The all--particle spectrum obtained
is shown for reference as well (dashed curve).  

Recent measurements in the 100~GeV region, i.e. AMS, BESS, CAPRICE, and IMAX,
obtain lower flux values as have been reported by earlier experiments.  These
values had not been available for an earlier compilation by Wiebel--Sooth et
al. \cite{wiebel}.Therefore, the flux in this reference is about 30\% larger
than the present result and the spectrum is slightly steeper.  

When taking the experimental data in Figure~\ref{directP} at face value the
reader might imagine a change of slope around $10^4$~GeV. However, the new
measurements do not confirm such former speculations.  Also indirect
measurements give no hint for deviations from a power law.  Deriving the
primary proton spectrum via the investigation of unaccompanied hadrons at
mountain level, the EAS--Top experiment finds no evidence for a structure in
the spectrum up to $10^5$~GeV \cite{castellina}.  

Figure~\ref{directHe} shows the differential energy spectrum for helium nuclei
as a function of the total particle energy.  The parameters for the spectrum,
obtained by a least--square fit, are
$\Phi_{He}^0=(5.71\pm0.09)\cdot10^{-2}$~(m$^2$\,s\,sr\,TeV)$^{-1}$ and
$\gamma_{He}=-2.64\pm0.02$ with $\chi^2\mbox{/d.o.f.}=3.5$.  The spectral index
agrees with the value obtained by Wiebel--Sooth et al., but the absolute flux
is now about 25\% lower due to contributions of recent measurements, both at
the low and high energy end of the spectrum.

As an example for heavy elements the energy spectrum for the most prominent
representative --- iron nuclei ---  is presented in Figure~\ref{directFe}.
Results from detectors which are able to resolve individual elements  are given
together with experiments, where the charge resolution allows only to specify
the flux for the iron group, e.g. JACEE, RUNJOB, Simon et al., and SOKOL.  The
best fit to all measurements shown above 1~TeV yields for the power law
$\Phi_{Fe}^0=(2.04\pm0.26)\cdot10^{-2}$~(m$^2$\,s\,sr\,TeV)$^{-1}$ and
$\gamma_{Fe}=-2.59\pm0.06$ with $\chi^2\mbox{/d.o.f.}=0.9$.  If the result is
compared to the earlier compilation, the present absolute flux is about 15\%
larger, while the spectral index agrees within the specified errors with the
old value.  The suppression of the flux at energies below 1~TeV is due to the
modulation in the heliosphere.  It can be inferred from the figure, that the
measurements are parametrized well using Equation~(\ref{solmod}).

\begin{figure}[bt]       
 \epsfig{file=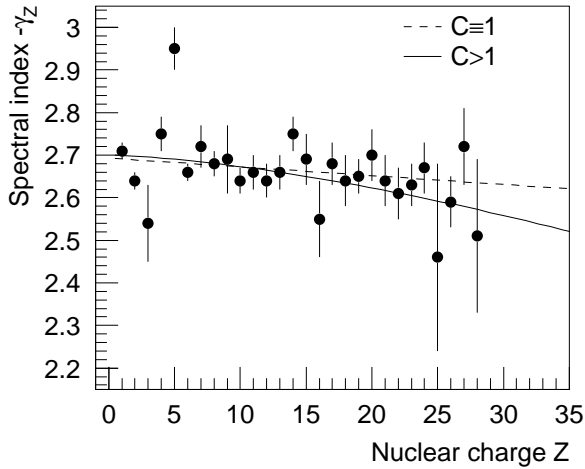,width=\columnwidth}
 \caption{\small Spectral index $\gamma_Z$ versus nuclear charge $Z$,
          see Table~\ref{directtab}. The solid line represents a three
	  parameter fit according to Equation~(\ref{gammafun}), the
	  dashed graph a linear fit.}
 \label{gamma}          
\end{figure}

\begin{table*}[hbt]\centering
\caption{\small Parameters of Equation~(\ref{gammafun}) for the linear and non--linear
         extrapolation of $\gamma_Z$.}
 \label{gammatab}
 \renewcommand{\arraystretch}{1.1}
\begin{tabular}{lccc}\hline
            & A            & B                           & C           \\ \hline
 linear     &$2.69\pm0.12$ &$(-2.07\pm1.05)\cdot10^{-3}$ & $\equiv1$   \\ 
 non--linear&$2.70\pm0.19$ &$(-8.34\pm4.67)\cdot10^{-4}$ &$1.51\pm0.13$\\ \hline
\end{tabular}
\end{table*}

The fit parameters for protons, helium, and iron nuclei are listed together
with results from Wiebel--Sooth et al.  for the elements hydrogen to nickel in
Table~\ref{directtab} in the appendix.  The spectral indices for these elements
are plotted in Figure~\ref{gamma} versus the nuclear charge.  The data show a
trend towards smaller values with increasing $Z$.  Such a $Z$--dependence of
$\gamma_Z$ could be explained by charge dependent effects in the acceleration
or propagation process.\\
Theories using a nonlinear model of Fermi acceleration in supernovae remnants
predict a more efficient acceleration for elements with a large mass to charge
ratio as compared to elements with a smaller $A/Z$ ratio, see e.g.
\cite{ellison}.  Consequently, it is expected that elements with higher $A/Z$
have a harder spectrum at the source. \\
The energy spectra observed at earth are modified during propagation of the
particles through the galaxy.  Some authors include reacceleration by weak
interstellar shocks in the standard leaky box model, e.g.
\cite{letaw,wandel,strong}.  Like the primary acceleration also the
reacceleration could be more efficient for high--$Z$--nuclei.

To estimate the fluxes of these ultra--heavy elements at high energies the
parametrization
\begin{equation}\label{gammafun}
 -\gamma_Z= A + B\cdot Z^C 
\end{equation}
is used to describe the $Z$ dependence of the spectral indices and to
extrapolate them to higher values.  To study systematic effects of the
extrapolation two approaches are used, a linear function ($C\equiv1$) and a
non--linear extrapolation, using all three parameters.  

The dashed line in Figure~\ref{gamma} represents the best fit of a linear
parametrization, exhibiting a decreasing spectral index as function of the
nuclear charge.  The data shown in the figure exhibit some curvature which
suggests to introduce the additional degree of freedom. If the parameter $C$ in
Equation~(\ref{gammafun}) is used as free parameter, the solid line in
Figure~\ref{gamma} is obtained.  The parameters for both trials are listed in
Table~\ref{gammatab}, both fits result in about the same
$\chi^2\mbox{/d.o.f.}\approx2.1$.  The values for the non--linear approach will
be corroborated below by an independent fit to the all--particle spectrum.

\subsection{Ultra--heavy elements}
\begin{figure}[hbt]
 \epsfig{file=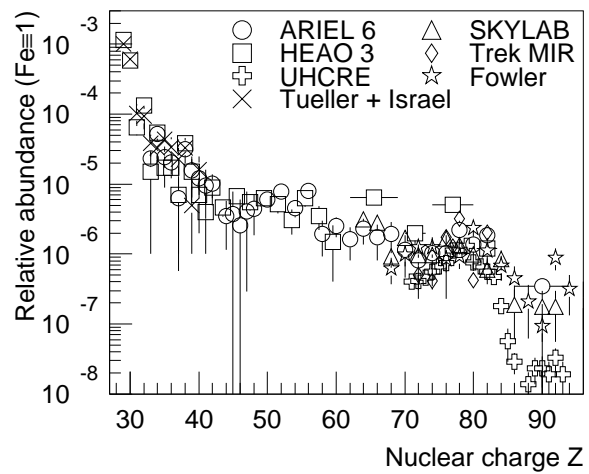,width=\columnwidth}
 \caption{\small Relative abundance of cosmic--ray elements ($Z>28$)
          normalized to Fe$\equiv 1$ from
          ARIEL~6 \cite{ariel},
          Fowler et al. \cite{fowler},
          HEAO 3 \cite{heao}, 
          SKYLAB \cite{skylab},
          TREK/MIR \cite{trek}, 
          Tueller et al. and Israel et al. \cite{tueller}, as well as
          UHCRE \cite{uhcre}.
	  The energy is around 1~GeV/nucleon.}
 \label{ariel}
\end{figure}
For ultra--heavy elements ($Z>28$) data exist only at relative low energies
around a few GeV/nucleon as already mentioned.  Figure~\ref{ariel} shows a
compilation of the relative abundance from copper $(Z=29)$ up to uranium
$(Z=92)$, as measured by several experiments on space crafts and balloons.  The
data are normalized to Fe$\equiv 1$, the threshold is about 0.5 to
1~GeV/nucleon.  Some authors give only results for groups of elements, this is
indicated by horizontal error bars.

The experiments
 ARIEL~6 \cite{ariel},
 HEAO 3 \cite{heao}, as well as
 Tueller et al. and Israel et al. \cite{tueller} 
quote abundances relative to iron.
Only relative abundances for elements $Z\ge70$ are reported by
 Fowler et al. \cite{fowler},
 SKYLAB \cite{skylab},
 TREK/MIR \cite{trek}, and
 UHCRE \cite{uhcre}. 
The results of the latter group have been normalized to ARIEL~6.  This detector
could resolve individual elements up to $Z=48$, and even charged nuclei above.
The range $70\le Z\le80$ has been used to match the abundances for Fowler et
al., SKYLAB, and UHCRE.  For TREK the interval $72\le Z\le80$ has been
utilized.

The results of all experiments show about the same structure for the relative
abundances.  For elements with $Z>80$ deviations are visible.  Due to the very
low flux only a few (\raisebox{-1.5mm}{$\stackrel{\textstyle<}{\sim}$}~10)
nuclei have been detected during a typical mission and the experiments reach
their limit for statistically reliable results.
 
The relative abundances shown in Figure~\ref{ariel} are the only data available
for ultra--heavy elements. No data about individual spectra are at disposal.
Hence, the spectral indices for these elements have to be estimated and are
obtained by extrapolating parametrization (\ref{gammafun}) to large $Z$.\\ 
The mean relative abundances for the elements are calculated from the data
given in Figure~\ref{ariel}, converted to absolute flux values and extrapolated
to higher energies in the following manner. The solar modulation is 
taken into account using Equation~(\ref{solmod}) and the fluxes are
extrapolated to 1~TeV/nucleus assuming at high energies power laws with the
extrapolated spectral indices.\\ 
For $Z>48$ the detectors could only resolve even charged elements, i.e. the
values shown are the sum of two elements. For the extrapolation they are split
between odd and even atomic numbers using the ratio 1:5. This ratio is a
typical value for the solar system composition, see Figure~\ref{flux}.  The
values thus obtained for $\Phi_Z^0$ and $\gamma_Z$, using the non--linear
extrapolation to obtain $\gamma_Z$, are listed in Table~\ref{directtab}.

\subsection{Comparison with solar system composition}
\begin{figure}[hbt]
 \epsfig{file=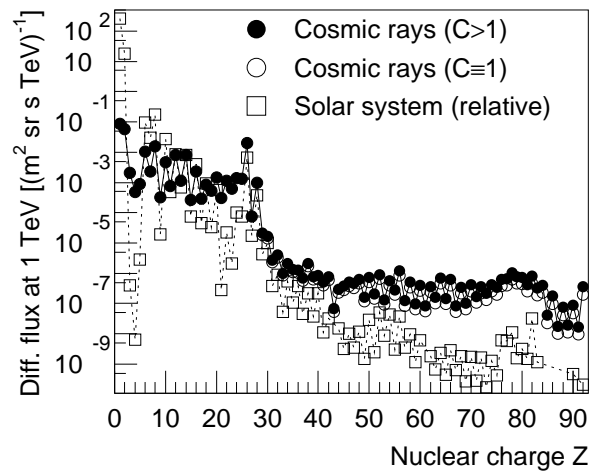,width=\columnwidth}
 \caption{\small Absolute flux $\Phi_Z^0$
          of cosmic--ray elements at $E_0=1$~TeV/nucleus 
          versus nuclear charge according to Table~\ref{directtab}.
	  Two versions for the extrapolation of $\gamma_Z$
	  for ultra--heavy nuclei are indicated, see text.
	  For comparison, solar system abundances are shown \cite{cameron}. 
	  The relative values for the solar
	  system have been normalized to the absolute flux of silicon ($Z=14$)
	  in cosmic rays. The ordinate accounts for cosmic rays only.}
 \label{flux}
\end{figure}

The absolute flux of all cosmic--ray elements at 1~TeV/nucleus, as listed in
Table~\ref{directtab}, is plotted in Figure~\ref{flux} versus the nuclear
charge. For the ultra--heavy elements the linear extrapolation ($C\equiv1$)
yields slightly smaller flux values as compared to the non--linear description
($C>1$).  For comparison, solar system abundances are shown as well. The
relative abundances for the solar system are normalized to the absolute flux of
silicon in cosmic rays.  The general structure of both distributions is about
the same and the differences between even and odd nuclear charge numbers are
mostly visible in both compositions.  However, there are differences which will
be discussed next. 

For cosmic rays the low abundance "valleys" in the solar system composition
around Z=4, 21, 46, and 70 are not present.  This is usually believed to be the
result of spallation of heavier nuclei during their propagation through the
galaxy.  Hydrogen, helium, and the CNO--group are suppressed in cosmic rays.
This has been explained by the high first ionization potential of these atoms
\cite{silberberg} or by the high volatility of these elements which do not
condense on interstellar grains \cite{meyer}.  Which property is the right
descriptor of cosmic--ray abundances has proved elusive, however, the
volatility seems to become the more accepted solution \cite{baring}.

\begin{figure*}\centering
 \epsfig{file=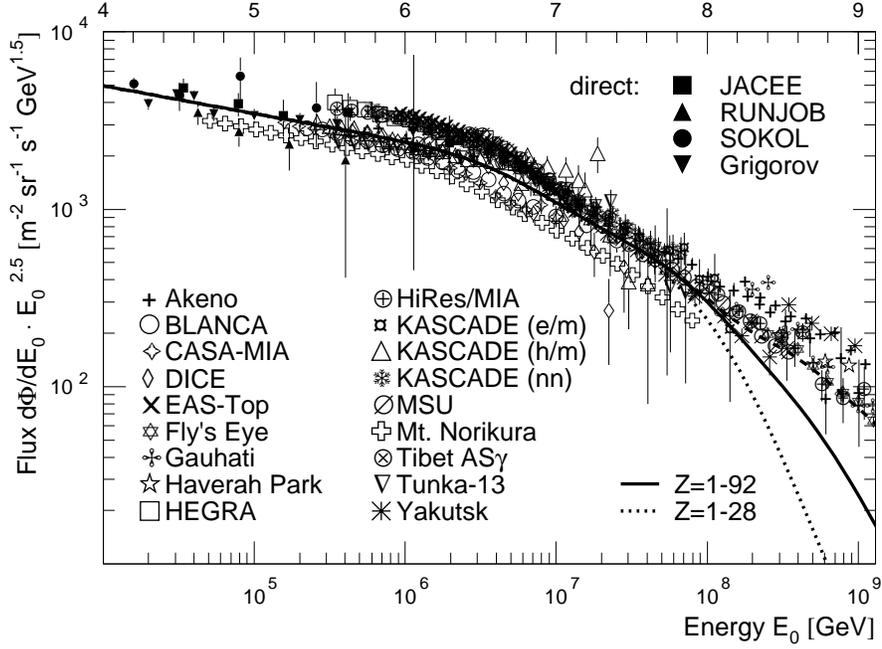,width=12cm}
 \caption{\small All--particle energy spectra obtained from direct and indirect
	  measurements, for references see Table~\ref{eshift} and text.  The
	  sum spectra for individual elements according to the \modell\ are
	  represented by the dotted line for $1\le Z\le28$ and by the solid
	  line for $1\le Z\le92$.  Above $10^8$~GeV the dashed line gives the
	  normalized average spectrum.}
 \label{knie0}
\end{figure*}

The iron group and the ultra--heavy elements are more pronounced in cosmic rays
as compared to the solar system.  Especially the r--process elements beyond
xenon ($Z$=54) are enhanced, partly due to spallation products of the platinum
and lead nuclei ($Z$=78, 82).  For the latter direct measurements at low
energies around 1~GeV/n yield about a factor two more abundance as compared to
the solar system and a factor of four for the actinides thorium and uranium
($Z$=90, 92) \cite{tsao}.  This has been attributed to the hypothesis that
cosmic rays are accelerated out of supernova ejecta--enriched matter
\cite{higdon}.

The differences for ultra--heavy elements between cosmic rays and solar system
abundances seen in Figure~\ref{flux} are much bigger as compared to the
measurements at energies of 1~GeV/nucleon. At low energies the flux is strongly
suppressed due to the solar modulation, while at 1~TeV/particle the effects of
the heliospheric magnetic fields are almost negligible.  Together with the
enhancements discussed before, this accounts for the differences seen in
Figure~\ref{flux}.  The overall distribution from hydrogen to uranium is much
flatter for cosmic rays.  While the solar system abundances cover a range of 11
decades, the cosmic--ray distribution extends over 6 decades in flux only.  

\section{Indirect Measurements}\label{indirect}

Many groups published results on the all--particle energy spectrum from
indirect measurements.  Several experiments detect the main components of
extensive air showers most commonly the electromagnetic component, but also
muons and hadrons are investigated.  The \v{C}erenkov photons produced by
relativistic shower particles and the fluorescence light of nitrogen molecules
in the atmosphere induced by air showers are also utilized.  The detectors are
located at various atmospheric depths corresponding to altitudes from 50~m up
to 4370~m a.s.l.. The experiments used in this compilation and their respective
measuring techniques as well as the atmospheric overburden are listed in
Table~\ref{eshift}.

\begin{table*}[hbt]
 \caption{\small Air shower experiments and components measured to derive the primary
   energy spectrum: e: electromagnetic, $\mu$: muonic, h: hadronic, \v{C}:
   \v{C}erenkov, and F: fluorescence light.  The particle thresholds are given
   for the muonic and hadronic components.  In addition, the atmospheric
   overburden [g/cm$^2$] and the shift of the energy scale are listed.}
 \label{eshift} 
 \begin{center}
 \renewcommand{\arraystretch}{1.1}
 \begin{tabular}{lcrrccrr}
  \hline 
  Experiment &e&\multicolumn{1}{c}{$\mu$}&
                \multicolumn{1}{c}{h}&\v{C}&F&
		\multicolumn{1}{c}{g/cm$^2$}&
		\multicolumn{1}{c}{Energy shift}\\
  \hline
  AKENO (low energy)  \cite{akeno}    &x&1~GeV  &      & & & 930&$- 4\%$\\
  BLANCA           \cite{blanca}      &x&       &      &x& & 870&$  4\%$\\
  CASA--MIA        \cite{casae}       &x&800~MeV&      & & & 870&$  4\%$\\
  DICE             \cite{dice}        &x&800~MeV&      &x& & 860&$  1\%$\\
  EAS--Top         \cite{eastop}      &x&1~GeV  &      & & & 820&$-11\%$\\
  HEGRA            \cite{hegra}       &x&       &      &x& & 790&$-10\%$\\
  KASCADE (electrons/muons) \cite{ulrich}  &x&230~MeV&      & & &1022&$- 7\%$\\
  KASCADE (hadrons/muons)  \cite{jrhknie} & &230~MeV&50~GeV& & &1022&$- 1\%$\\
  KASCADE (neural network) \cite{roth}    &x&230~MeV&      & & &1022&$- 8\%$\\
  MSU              \cite{msu}         &x&       &      & & &1020&$- 5\%$\\
  Mt. Norikura     \cite{norikura}    &x&       &      & & & 735&$  9\%$\\
  Tibet            \cite{tibet}       &x&       &      & & & 606&$-10\%$\\
  Tunka--13        \cite{tunka}       & &       &      &x& & 680&$  0\%$\\
  Yakutsk (low energy) \cite{yakutsk} & &       &      &x& &1020&$- 3\%$\\
  &&&&&&&\\
  AKENO (high energy) \cite{akenohe}  &x&       &      & & & 930&$-16\%$\\
  Fly's Eye        \cite{flyseye}     & &       &      & &x& 860&$- 3\%$\\
  Gauhati          \cite{gauhati}     &x&       &      & & &1025&$- 5\%$\\
  Haverah Park     \cite{haverahpark} &x&       &      & & &1018&$-10\%$\\
  HiRes--MIA       \cite{hires}       & &800~MeV&      & &x& 860&$- 5\%$\\
  Yakutsk (high energy) \cite{yakutskhe}&x&1~GeV&      &x& &1020&$-20\%$\\
  \hline 
  \end{tabular}
  \end{center}
\end{table*}

\begin{table*}
 \caption{\small Spectral index around 400~PeV measured by three experiments, 
          from \cite{flyseye}.}
 \label{diptab}
 \begin{center}
 \renewcommand{\arraystretch}{1.1}
 \begin{tabular}{lcccccc}\hline
  Experiment   &&energy range&spectral index&&energy range&spectral index\\
  \cline{1-1}\cline{3-4}\cline{6-7}
  AKENO        && $10^{15.7}-10^{17.8}$ :& $3.02\pm0.03$ && 
                 $10^{17.8}-10^{18.8}$ :& $3.16\pm0.08$ \\
  Haverah Park && $10^{17.48}-10^{17.6}$:& $3.01\pm0.02$ && 
                 $10^{17.6}-10^{18.6}$ :& $3.24\pm0.07$ \\
  Fly's Eye    && $10^{17.3}-10^{17.6}$ :& $3.01\pm0.06$ && 
                 $10^{17.6}-10^{18.5}$ :& $3.27\pm0.02$ \\ \hline
 \end{tabular}
 \end{center}
\end{table*}

Some of the experiments present different results, based on different hadronic
interaction models used to interpret the data.  Detailed studies of air shower
properties, especially concerning the hadronic component, have been performed
by the KASCADE group \cite{wwpap,jens}.  Comparison to simulations, using the
program CORSIKA \cite{corsika} with several high--energy hadronic interaction
models implemented revealed, that the QGSJET model presently is the best model
to describe air shower data.  Other groups obtained similar results, see e.g.
\cite{ewqgsjet,blanca}. Hence, if more than one interpretation of measurements
is given, the energy spectrum derived with the QGSJET model is used.

The results of the groups are presented in Figure~\ref{knie0}.  The overall
agreement between the experiments is quite good, the differential fluxes
multiplied by $E^{2.5}$ agree within a factor of two.  All experiments exhibit
a similar shape of the spectrum despite their different absolute normalization.
This is remarkable since different components of air showers are investigated
by various groups at different atmospheric depths, i.e.  different types of
particle interactions in the atmosphere are probed and the showers are sampled
at different stages of their longitudinal development.

The all--particle spectrum obtained by direct measurements from
JACEE \cite{jaceefe},
RUNJOB \cite{runjoball},
SOKOL \cite{sokol}, and
Grigorov et al. \cite{grigorov}
agrees with the indirect measurements in the region of overlap, i.e.
from below 100~TeV to 1~PeV particle energy.

\begin{figure}[hbt]\centering
 \epsfig{file=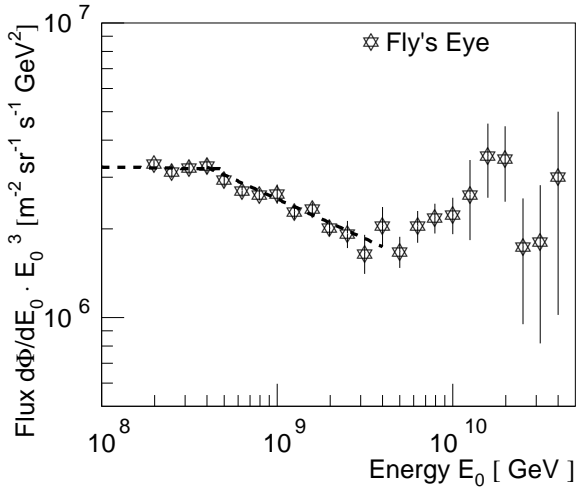,width=\columnwidth}
 \caption{\small All--particle energy spectrum at very high energies according to
          the Fly's Eye experiment.
	  The dashed line represents a fit by Bird et al. \cite{flyseye}.}
 \label{flyseye0}
\end{figure}

The {\sl knee} at about 4~PeV is clearly recognizable in the spectrum.
Assuming this bend being caused by the cut--off of the proton component, the
galactic component extends in the \modell\ up to $Z_U\cdot 4$~PeV $\approx
0.4$~EeV.  The energy spectrum multiplied by $E_0^3$ is shown in
Figure~\ref{flyseye0} for high energies as reported by the Fly's Eye group.
Indeed, a change in the spectral slope around $4 \cdot 10^8$ GeV is visible,
the dashed line represents a fit taken from the Fly's Eye publication
\cite{flyseye}.  A similar structure has been observed by the experiments AKENO
and Haverah Park, the spectral indices obtained around 400~PeV by the three
experiments are summarized in Table~\ref{diptab}.  The change coincides well
with the anticipated cut--off of the heaviest nuclei of the galactic component.

\section{Cosmic--ray energy spectrum} \label{espec}
\begin{figure*}[hbt]\centering
 \epsfig{file=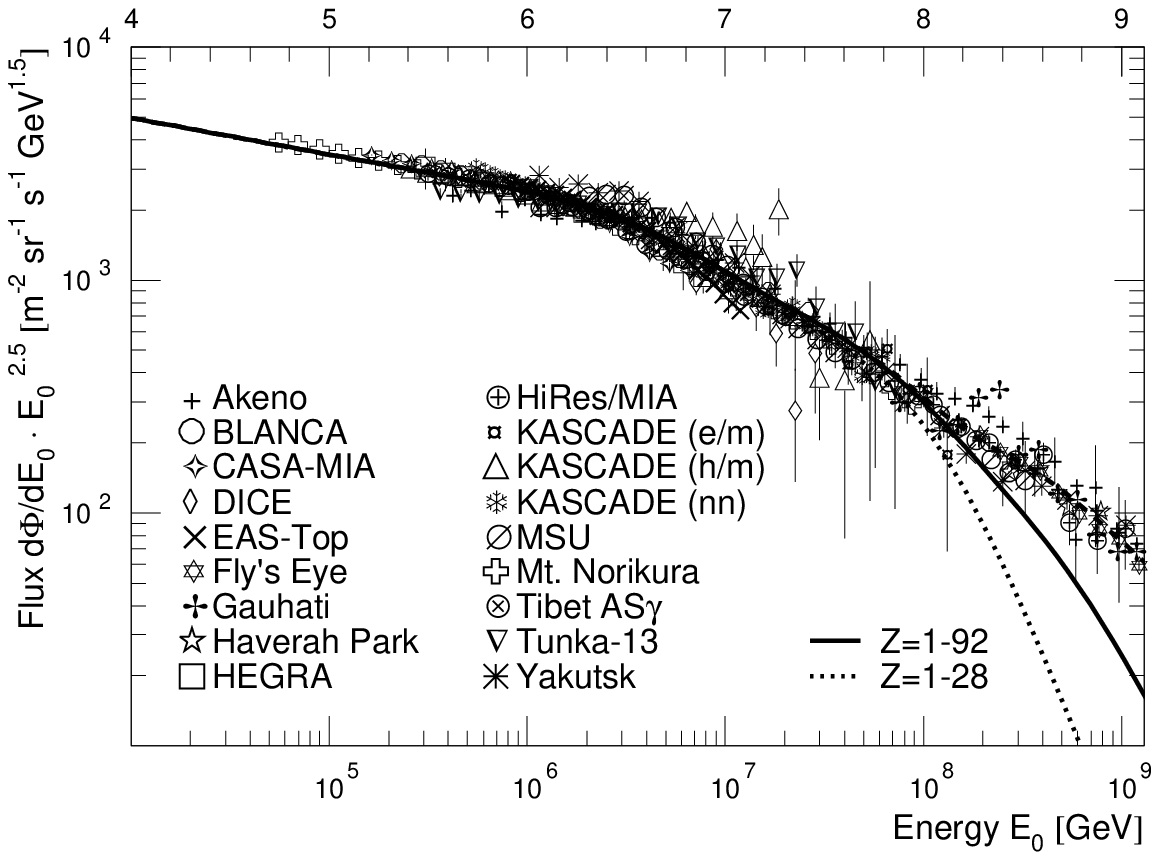,width=12cm}
 \caption{\small Normalized all--particle energy spectra for individual experiments. 
	  The renormalization values for the energy scale and references are
	  given in Table~\ref{eshift}.  The sum spectra for individual elements
	  according to the \modell\ are represented by the dotted line for
	  $1\le Z\le28$ and by the solid line for $1\le Z\le92$.  Above
	  $10^8$~GeV the dashed line reflects the average spectrum.}
 \label{knie1}
\end{figure*}

The different absolute normalizations between the experiments, apparent 
in Figure~\ref{knie0}, are probably caused by uncertainties in the energy
calibration. The latter are quoted by the experiments to be in the order of 
10\% to 20\%. That means the energy scales may be shifted by this amount
with respect to each other. 
This has been done in order to elaborate a more consistent all--experiment
spectrum.

\subsection{Renormalization}
Data from experiments which start to measure below 1~PeV have been renormalized
to fit the all--particle flux at 1~PeV extrapolated from the direct
measurements.  The solid line in Figure~\ref{knie0} below 1~PeV represents the
average flux of these measurements.  The renormalization factors obtained are
listed in Table~\ref{eshift} for the detector stations AKENO (low energy) to
Yakutsk (low energy).  Only small shifts $(\le11\%)$ are necessary, all within
the mentioned energy uncertainties. The mean value amounts to $(-2.9\pm1.7)\%$.

For the experiments measuring at very high energies (AKENO (high energy) to
Yakutsk (high energy) in Table~\ref{eshift}) the energy scale has been checked
subsequently in the following way.  The spectra have been arranged  according
to the lower limit of the given energy range, i.e. HiRes, AKENO, Fly's Eye,
Haverah Park, and Yakutsk, and then normalized to the average flux of all
previous experiments at their lowest energies.  The shape of the Gauhati energy
spectrum is systematically different from all others. Therefore, it has been
normalized to fit the average flux obtained by all other experiments above
$10^8$~GeV.  The energy shifts necessary are listed in Table~\ref{eshift}, they
are all negative with an average value $(-9.8\pm2.8)\%$.

Due to differences in the shape of the spectra of individual experiments with
respect to the average spectrum it is not always obvious      how to normalize.
This yields errors of about $\pm2\%$ for the renormalization of the energy
scales given.  The average energy corrections in both groups are negative and
average to $(-5.0\pm1.6)\%$. This indicates that the experiments tend to
overestimate the energy, which could be explained by a systematic effect in the
hadronic interaction models applied.

The normalized spectra are shown in Figure~\ref{knie1}. The spread between the
different results has shrunk, all fluxes are contained in a narrow band.  The
similar shape becomes obvious in this representation, an interesting
observation, since the energy scaling does not change the shape of the spectra.

The weighted average of the normalized all-particle spectra has been
calculated, taking the errors of the individual measurements into account, the
result is presented in Figure~\ref{meanspec0} and Table~\ref{etab}.  The errors
reflect the r.m.s. values of the data. The shape of this multi--experiment
all--particle energy spectrum, already seen in Figure~\ref{knie1}, becomes
clearly apparent in this plot.

\subsection{Energy dependence for the cut--off}\label{edependence}

\begin{figure*}[hbt]\centering
 \epsfig{file=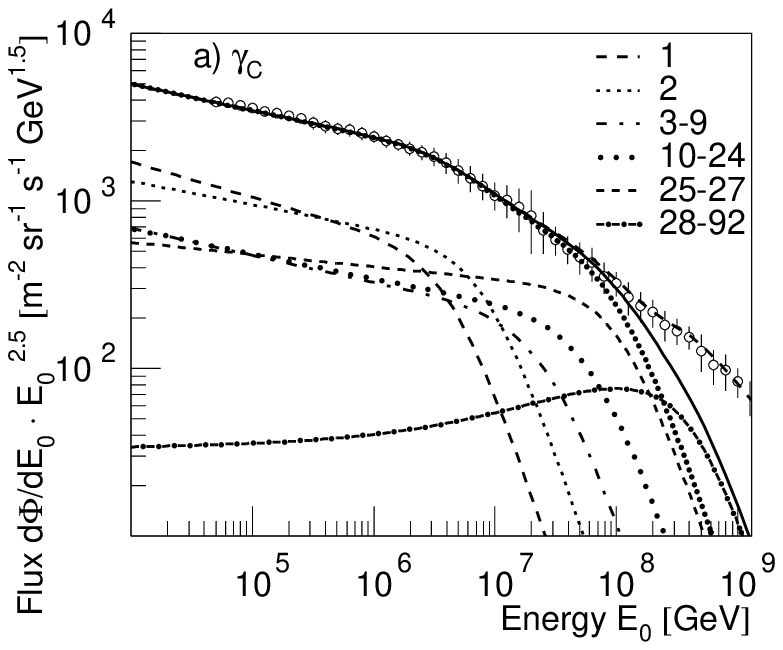,width=\columnwidth}
 \epsfig{file=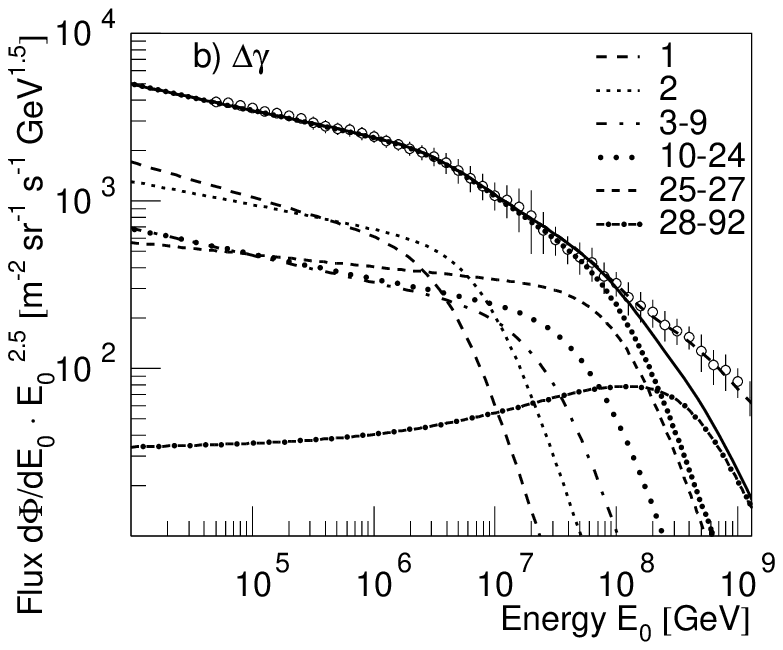,width=\columnwidth}
 \caption{\small Average all--particle energy spectrum. 
	  The line through the data represents a fit of the sum spectrum for
	  elements with $1\le Z\le92$ according to the \modell\ with rigidity
	  dependent cut--off for a)~common $\gamma_c$ and b)~common
	  $\Delta\gamma$. The dotted line shows the spectrum for $1\le Z\le28$.
	  In addition, energy spectra for groups of elements are shown.  Above
	  $10^8$~GeV the dashed line reflects the average spectrum.}
 \label{meanspec0}
\end{figure*}

\begin{table*}
 \caption{\small Parameters to fit the all--particle energy spectrum 
          for the different approaches to describe the cut--off
	  behaviour, see Equations~(\ref{specfun}) to (\ref{cutofffun}), 
	  with the \modell.}
 \label{fitrestab}$$
 \renewcommand{\arraystretch}{1.1}
 \begin{array}{rcccccl} \hline
  \mbox{cut--off:}  &\mbox{rigidity} &~~~&\mbox{mass} &~~~& \mbox{constant}&\\
                    &\mbox{dependent}&&\mbox{dependent}&&                &\\
  \hat{E}_Z=             & \hat{E}_p\cdot Z&&\hat{E}_p\cdot A  && \hat{E}_p    &\\ \hline
  \hat{E}_p~\mbox{[PeV]}=& 4.51\pm0.52     &&3.66\pm0.41 && 3.50\pm0.38 &\mbox{common~} \gamma_c\\
  \gamma_c=          &-4.68\pm0.23    &&-7.82\pm1.09 &&-3.06\pm0.02 &\\
  \epsilon_c=        & 1.87\pm0.18    && 2.30\pm0.23 && 1.94\pm0.51 &\\
  \chi^2\mbox{/d.o.f}=& 0.116          && 0.290       && 0.086       &\\ \hline
  \hat{E}_p~\mbox{[PeV]}=& 4.49\pm0.51    && 3.81\pm0.43 && 3.68\pm0.39 &\mbox{common~} \Delta\gamma\\
  \Delta\gamma=      & 2.10\pm0.24    && 5.70\pm1.23 && 0.44\pm0.02 &\\
  \epsilon_c=        & 1.90\pm0.19    && 2.32\pm0.22 && 1.84\pm0.45 &\\
  \chi^2\mbox{/d.o.f.}=& 0.113        && 0.292       && 0.088       &\\ \hline
 \end{array}$$
\end{table*}

The all--particle spectrum is used to determine the three parameters of the
\modell.  A least--square fit uses the values $\Phi_Z^0$ and $\gamma_Z$, as
given in Table~\ref{directtab}, and assumes a rigidity dependent cut--off.  The
two versions to model the cut--off behaviour are investigated, viz. a common
spectral index $\gamma_c$ above the {\sl knee} for all elements ---
Equation~(\ref{specfun}) --- or a common difference in spectral slope
$\Delta\gamma$ --- Equation~(\ref{specfundg}).  The best fits obtained are
shown in Figure~\ref{meanspec0} as solid lines for a common $\gamma_c$
(Figure~\ref{meanspec0}\,a) and for a common $\Delta\gamma$
(Figure~\ref{meanspec0}\,b).  The parameters obtained are listed in the first
column of Table~\ref{fitrestab}.  

As seen in the figure, both approaches describe the data quite well and yield
low $\chi^2$/d.o.f. values which are given in the table. The r.m.s. values of
the data have been used as errors for the fit procedure, thus the $\chi^2$
values obtained are $<1$.  But their relative numbers are still a good quantity
to characterize the different fit results. The description of the
cut--off behaviour according to Equations~(\ref{specfun}) or (\ref{specfundg})
is not crucial for the results obtained. Both possibilities yield almost the
same shape of the energy spectra as can be seen in Figure~\ref{meanspec0}. To
check systematic effects, different bin sizes for the all--particle spectrum
have been tried and the upper limit of the fit range has been varied from 20 to
150~PeV.  All fit results agree within the quoted errors. 

To derive the energy spectra for ultra--heavy elements, the non--linear
extrapolation for $\gamma_Z$ with $C>1$ has been used. Applying the linear
extrapolation results in a sum spectrum which nearly coincides with the dotted
line in Figure~\ref{meanspec0} representing the sum spectra for $1\le Z\le28$.
The all--particle spectrum below $10^8$~GeV is not effected by the choice for
the extrapolation. Above this energy the contribution of the sum spectrum to
the all--particle spectrum for $1\le Z\le92$ is larger for the non--linear
extrapolation. For example, at $5\cdot10^8$~GeV the sum spectrum contributes
with about 50\% for the non--linear and with 12\% for the linear extrapolation.

A consistency check concerning the $\gamma_Z$ extrapolation has been performed
in the following way: The sum spectrum for $1\le Z\le 92$ has been fitted to
the all--particle spectrum and the spectral indices for $Z>28$ have been
obtained from this fit, i.e. the parameters $\hat{E}_p$, $\epsilon_c$,
$\gamma_c$ or $\Delta\gamma$ in Equations~(\ref{specfun}) or (\ref{specfundg})
and the parameters $A$,$B$, and $C$ in Equation~(\ref{gammafun}) have been
determined simultaneously in a  six--parameter fit.  The values for both
methods ($\gamma_c$ and $\Delta\gamma$) are consistent with the results listed
in Table~\ref{fitrestab}. The parameters for $\gamma_Z$ are almost identical to
the values for the non--linear extrapolation given in Table~\ref{gammatab}.
For this six--parameter fit the upper fit boundary has been varied from
30~PeV to 500~PeV. It is worth mentioning that the fit can not lower the
exponents of the ultra--heavy elements even more in order to fill up the
missing flux to the all--particle spectrum in the energy region above 600~PeV.
The boundary of the all--particle spectrum at lower energies is the limiting
factor.\\
That means, the values for $\gamma_Z$ have been derived in two independent
ways, a fit of $\gamma_Z$ versus $Z$ in Figure~\ref{gamma} and in a fit of the
all--particle spectrum with the \modell.  Both fits yield the same results, an
intriguing observation, favouring the non--linear extrapolation of $\gamma_Z$.

\begin{figure*}[hbt]\centering
 \epsfig{file=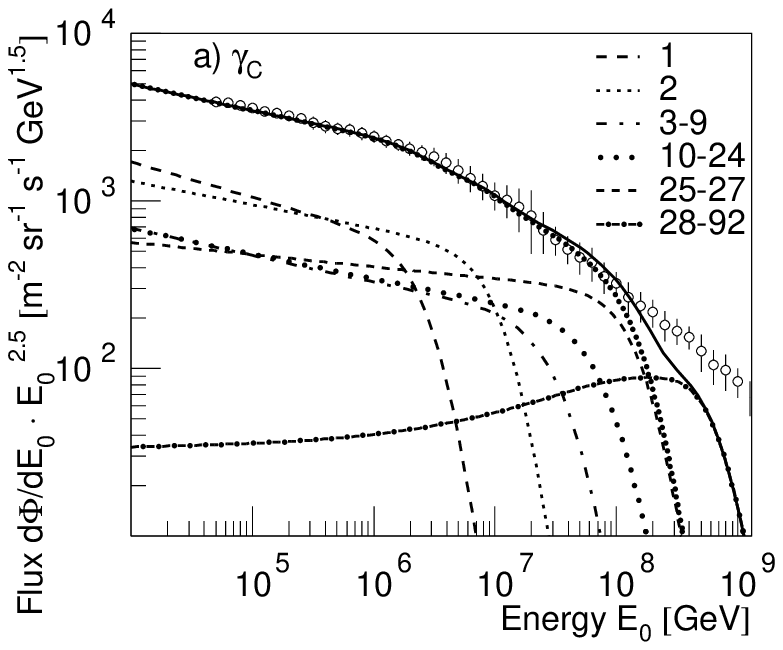,width=\columnwidth}
 \epsfig{file=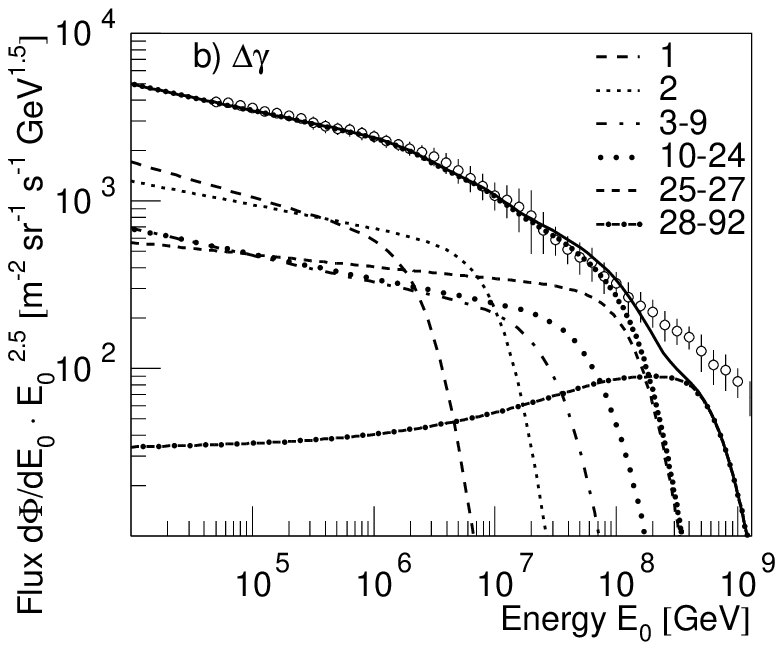,width=\columnwidth}
 \caption{\small Average all--particle energy spectrum. 
	  The line through the data represents a fit of the sum spectrum for
	  elements with $1\le Z\le92$ according to the \modell\ with mass
	  dependent cut--off for a)~common $\gamma_c$ and b)~common
	  $\Delta\gamma$. The dotted line shows the spectrum for $1\le Z\le28$.
	  In addition, energy spectra for groups of elements are shown.}
 \label{meanspec1}
\end{figure*}

All parameters of Equation~(\ref{specfun}) and (\ref{specfundg}) being
determined, the spectra of individual elements can be calculated.  In order to
avoid confusion not all individual spectra, but only graphs for six groups of
elements are plotted in Figure~\ref{meanspec0} for demonstration.  One observes
that around 50~PeV about 50\% of the particles should belong to the iron group
and hydrogen contributes only with a few \% to the all--particle flux.  Indeed,
the AKENO experiment \cite{hara} found when determining the mass composition
from electron and muon sizes above 20~PeV that the composition becomes heavier
with increasing energy.  In an analysis with three mass groups (proton--helium,
CNO, and iron) the amount of the light group turned out to be small, viz. in
the 10\% region or below around 50~PeV.

The average spectrum shows some slight structure above 4.5~PeV, i.e. it is
not a pure power law. In the phenomenological approach such a shape is
attributed to the sum of 92 spectra with individual cut--off energies.

Some authors, see e.g. \cite{terantonian}, propose an additional component
being necessary in addition to the known fluxes from direct observations in
order to describe the all--particle energy spectrum in the {\sl knee} region.
It may be pointed out, that the sum spectrum of all individual elements is
sufficient to compose the all--particle spectrum up to 100~PeV, as can be
inferred from Figures~\ref{knie1} and \ref{meanspec0}.  No additional
cosmic--ray component is required in the {\sl knee} region to explain the
observed spectrum.

In order to check the two other assumptions about the cut--off behaviour of the
individual element spectra --- see Equation~(\ref{cutofffun}) --- the average
energy spectrum has been fitted using these two hypotheses as well.

The results for the mass dependent cut--off using Equations~(\ref{specfun}) and
(\ref{specfundg}) are shown in Figure~\ref{meanspec1} and the fit values
obtained are listed in the second column of Table~\ref{fitrestab}. Again, the
resulting spectra are essentially independent of the ansatz used to describe
the spectral indices above the cut--off.  The upper end of the fit range has
been varied from 50 to 200~PeV, yielding consistent results. 

The sum spectra obtained describe the overall shape of the spectrum less well
as compared to the rigidity dependent assumption. They are below the
experimental data around 3~PeV and above them around 50~PeV, this is reflected
in a worse $\chi^2$/d.o.f. value of about a factor of 2.5 when compared to the
rigidity dependent choice. Also, the steepening around the individual {\sl
knees} turns out to be quite different when comparing the spectra of the
elemental groups in Figures~\ref{meanspec0} and \ref{meanspec1}.  For the
rigidity dependent cut--off the fit yields a moderate steepening, viz. a change
of the slope of $\Delta\gamma= 2.10$ or an index of $\gamma_c=-4.7$ and a
smooth transition between the power laws, ranging across slightly less than one
decade, with $\epsilon=1.9$. \\ 
The mass dependent ansatz results in a strong steepening of the spectra to
$\gamma_c=-7.8$ above the {\sl knee} or a change of the spectral index of
$\Delta\gamma= 5.7$ within a region of about half a decade ($\epsilon=2.3$).
Taking the propagation of cosmic rays through the galaxy into consideration,
such a sharp cut--off would be hard to explain on astrophysical grounds.  Maybe
a nearby cosmic--ray source, as discussed in \cite{wolfendale}, or a new type
of interaction in the atmosphere could lead to such a sharp cut--off.

\begin{figure*}[hbt]\centering
 \epsfig{file=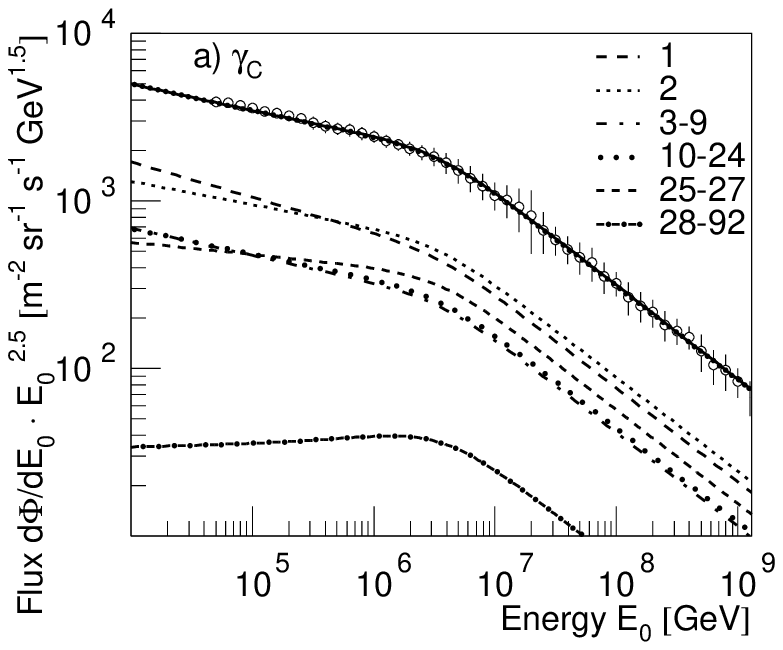,width=\columnwidth}
 \epsfig{file=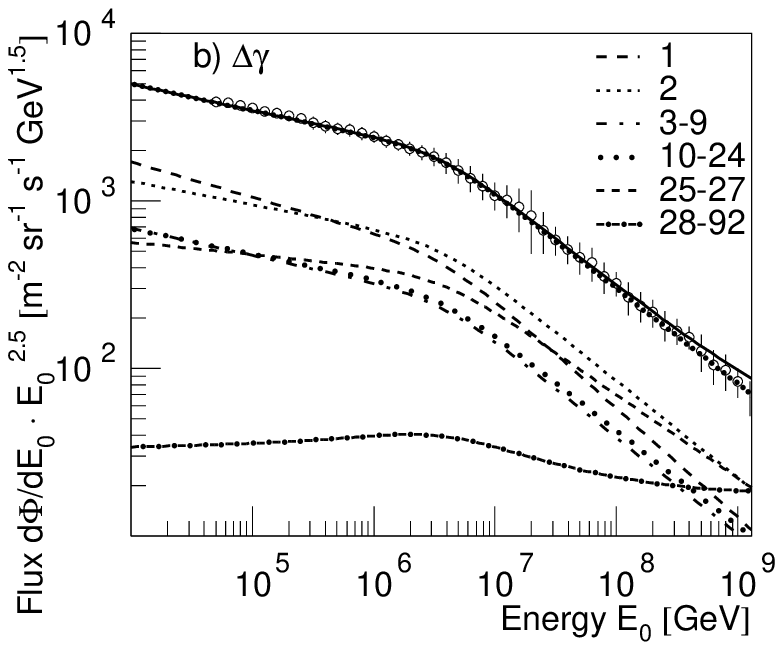,width=\columnwidth}
 \caption{\small Average all--particle energy spectrum. 
	  The line through the data represents a fit of the sum spectrum for
	  elements with $1\le Z\le92$ according to the \modell\ with constant
	  cut--off energy for a)~common $\gamma_c$ and b)~common
	  $\Delta\gamma$.  In addition, energy spectra for groups of elements
	  are shown.}
 \label{meanspec2}
\end{figure*}

Using the linear extrapolation for $\gamma_Z$ does not improve the
$\chi^2$--value, the spectrum around 3~PeV can not be described satisfactorily
by mass--dependent cut--off energies.  Nevertheless, a recent analysis,
combining electromagnetic shower size spectra of several experiments favours a
{\sl knee} for individual elements proportional to their mass \cite{schatz}.

\begin{figure}[hbt]\centering
 \epsfig{file=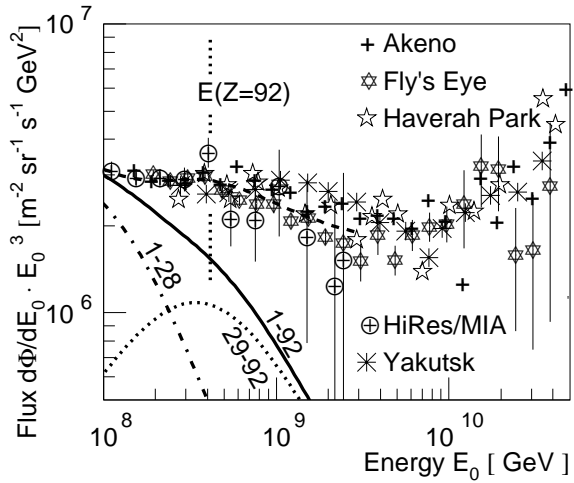,width=\columnwidth}
 \caption{\small Normalized all--particle energy spectra multiplied by $E_0^3$ for
     several experiments. For normalization factors and references see
     Table~\ref{eshift}.  The dashed line represents the average flux.  The sum
     spectrum according to the \modell\ is represented by the solid line for
     all elements ($Z$=1--92), as well as for two elemental groups.  The
     cut--off energy $\hat{E}_U=414$~PeV is marked by the pointed vertical
     line.}
 \label{flyseye1}
\end{figure}

When probing the third hypothesis for the cut--off energy, the constant
cut--off, the upper fit boundary has been changed from 20 to 560~PeV resulting
in consistent parameters, too. The obtained spectra are plotted in
Figure~\ref{meanspec2} for the two assumptions on the spectral indices above
the {\sl knee} and the corresponding parameters are listed in the last column
of Table~\ref{fitrestab}.  For the common $\gamma_c$ the differences between
the linear and the non--linear extrapolation for $\gamma_Z$ are
negligible.  For a common $\Delta\gamma$ deviations above 200~PeV are
significant, as can be inferred from Figure~\ref{meanspec2}\,b.
The non--linear extrapolation (solid line) results in larger flux values
as compared to the linear extrapolation which coincides nearly with the
sum spectrum for $1\le Z\le28$ (dotted line).\\
The shape of the average energy spectrum is well described up to 200~PeV, as
indicated by the low $\chi^2$/d.o.f. values. But the change in the spectrum at
400~PeV, seen in Figures~\ref{flyseye0} and \ref{flyseye1}, can not be
described. The all--particle spectrum lies significantly above the measured
values.  A very smooth transition is obtained, the spectral indices change only
little by $\Delta\gamma=0.4$ to a value of $\gamma_c=-3.1$ above the {\sl
knee}.  The width of the transition region, described by $\epsilon=1.9$ is
comparable to the one for the rigidity dependent cut--off.  However, the ansatz
of a constant cut--off will be disfavoured below, when discussing the resulting
mean logarithmic mass. 

Combining all arguments discussed, the rigidity dependent cut--off is the best
choice to describe the all--particle spectrum.  In the following the
non--linear extrapolation for $\gamma_Z$ will be used together with the
hypothesis of a common $\Delta\gamma$ as shown in Figure~\ref{meanspec0}\,b.

The resulting spectra with a rigidity dependent cut--off for protons, He, and
Fe are represented in Figures~\ref{directP} to \ref{directFe} as solid lines
with a dotted continuation in the {\sl knee} region.  The all--particle flux
summed up from all individual elements is shown as dashed line for reference.
In Figures~\ref{knie0} and \ref{knie1} the sum spectrum of individual elements
is given for $1\le Z\le92$ as solid line and as dotted line for $1\le Z\le28$.

The normalized spectra for experiments investigating very high energies are
shown in Figure~\ref{flyseye1} with the flux being multiplied by $E_0^3$.  The
slight steepening in the spectral slope around 0.4~EeV can be recognized in
the results of AKENO, Fly's Eye, Haverah Park, and HiRes/MIA. A similar
steepening occurs in the Yakutsk data, but at about 4 times higher energies.
Due to this unexplained systematic effect, the Yakutsk data are not taken into
account when calculating the average flux.  The latter is represented by a
dashed line up to 3~EeV.  Sum spectra for two groups of elements using a
rigidity dependent cut--off with a common $\Delta\gamma$, as presented in
Figure~\ref{meanspec0}\,b, are shown and it can be inferred, that the steepening
of the spectrum coincides with the cut--off of the heaviest nuclei according to
the \modell. The energy $\hat{E}_U=92\cdot \hat{E}_p=414$~PeV is marked for
reference. 

\begin{figure}[hbt]\centering
 \epsfig{file=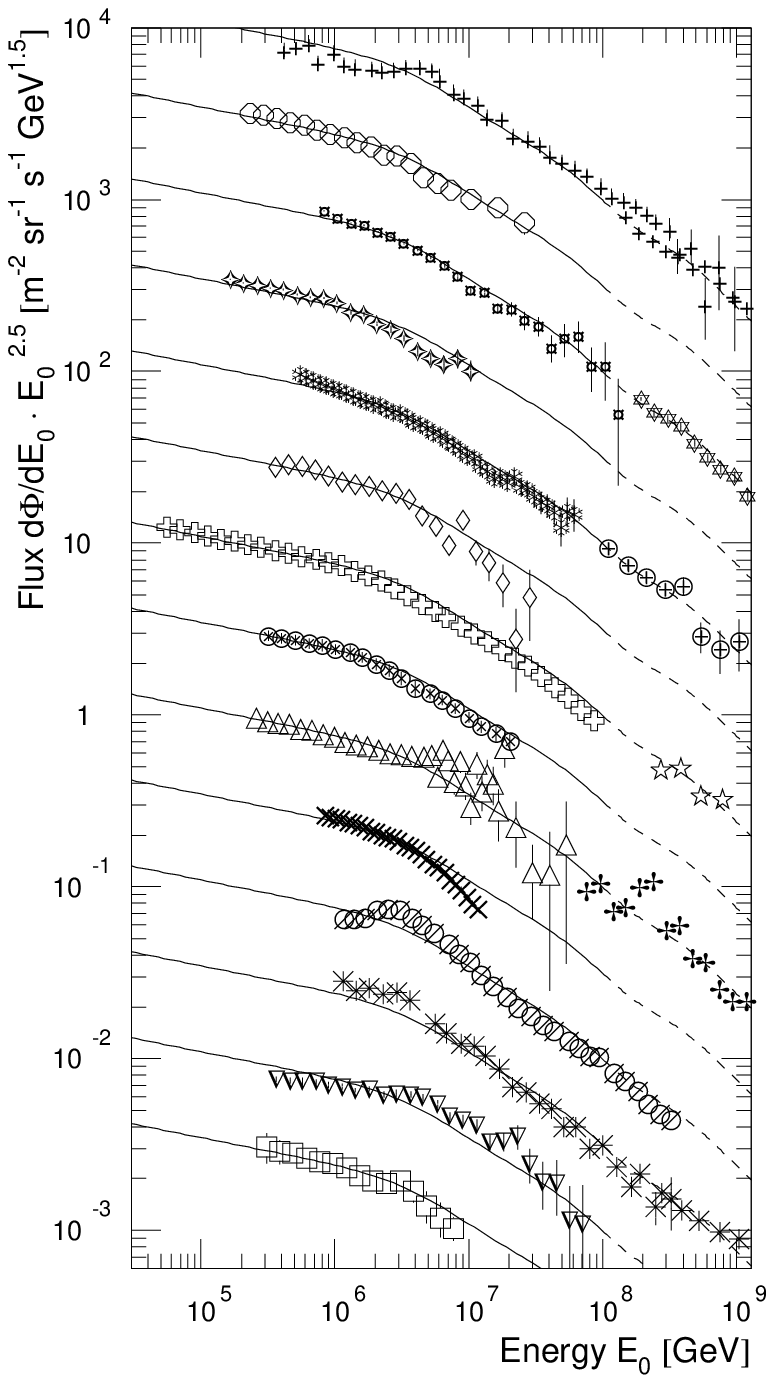,width=\columnwidth}
 \epsfig{file=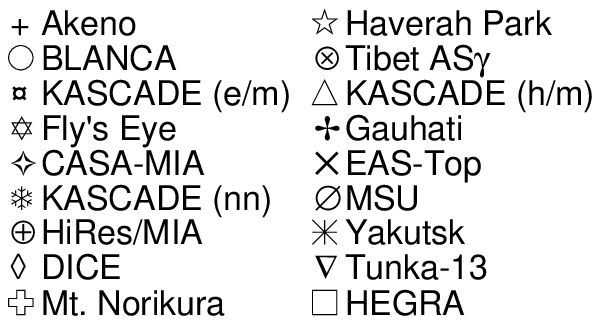,width=0.9\columnwidth,
         clip=,bbllx=37, bblly=46,bburx=233, bbury=143} 
 \caption{\small Normalized all--particle energy spectra for individual experiments 
          compared to the all--particle spectrum of the \modell.
	  The individual results are shifted in steps of half a decade in flux
	  in order to reduce overlap.}
 \label{knie3}
\end{figure}

This may lead to the conjecture, that the second {\sl knee} in the
all--particle spectrum is caused by the end of the galactic component of
cosmic--rays, i.e. by the cut--off of the heaviest elements.  At higher
energies an additional component is required to account for the observed flux.
For example, at 400~PeV the sum for $1\le Z\le92$ accounts for about 55\% of
all particles.  The linear extrapolation for $\gamma_Z$ yields a spectrum which
nearly coincides with the spectrum shown for $1\le Z\le 28$.
Figure~\ref{flyseye1} shows, that ultra--heavy elements with the non--linear
dependence of $\gamma_Z$ on $Z$ are important to describe the measured
all--particle spectrum in the region from 0.1 to 1~EeV and to explain the
second {\sl knee}.

In the figures presented so far it was hard to recognize the shape of the
individual spectra with respect to each other.  To elucidate the situation, the
normalized results of the individual experiments are compared in
Figure~\ref{knie3} to the all--particle spectrum.  The solid lines represent
the sum spectrum up to $10^8$~GeV, the dashed lines above represent the average
spectrum.  In order to reduce overlap between the graphs, the spectra have been
shifted in flux in steps of half a decade.  Different air shower components,
including electrons, muons, hadrons, \v{C}erenkov, and fluorescence photons,
have been used to derive the spectra, see Table~\ref{eshift}, and the results
are compatible with the average all--particle spectrum.  However, for some
experiments systematic discrepancies to the all--particle spectrum can be
recognized.

As the final result of the present investigations   the all-particle energy
spectrum derived according to the \modell\ with a rigidity dependent cut--off
and a common $\Delta\gamma$ is given in Table~\ref{etab} for reference.  The
differences between the calculated spectrum and the average measured flux ---
see Figure~\ref{meanspec0}\,b --- are given as errors.  The measurements start
at 0.5~PeV in the figure, consequently, below this energy no errors are quoted.
The mean relative deviation of the calculated spectrum amounts only to
remarkable~3\%.

\section{Cosmic--ray mass composition} \label{secmasse}
\begin{figure*}[hbt]\centering
 \epsfig{file=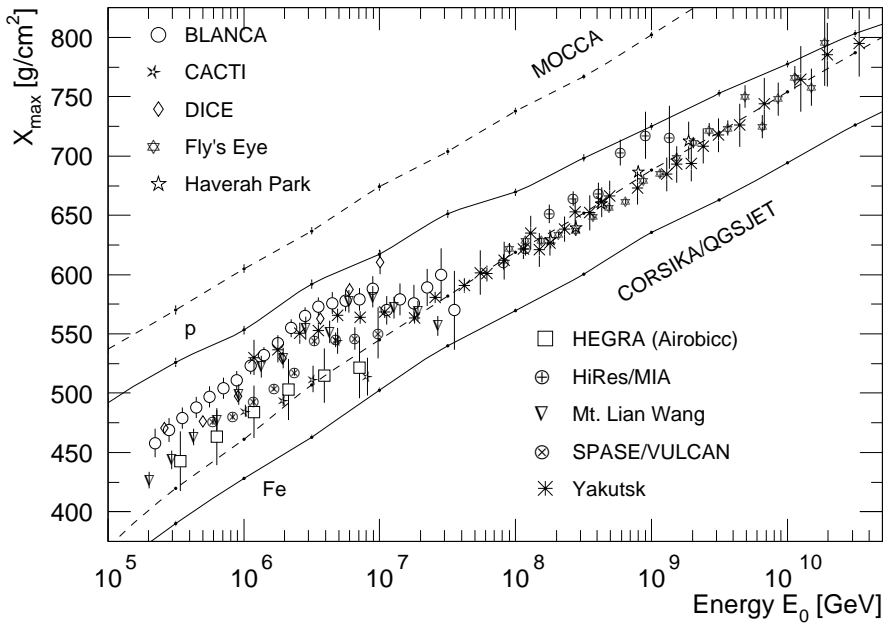,width=12cm}
 \caption{\small Average depth of the shower maximum $X_{max}$ versus primary 
	  energy as obtained by different experiments, for references see
	  Table~\ref{massetab}.  Simulations taken from \cite{heck,pryke} using
	  CORSIKA/QGSJET (solid lines) and MOCCA (dashed lines) are given for
	  primary protons and iron nuclei.}
 \label{xmax}
\end{figure*}

An often--used quantity to characterize the cosmic--ray mass composition above
1~PeV is the mean logarithmic mass, defined as 
\begin{equation}\label{lna} \lna= \sum_i r_i \ln A_i \quad, \end{equation} 
$r_i$ being the relative
fraction of nuclei of mass $A_i$.  Some experiments derive this observable
under certain assumptions from their measurements.  In the superposition model
of air showers, the shower development of heavy nuclei with mass $A$ and energy
$E_0$ is described by the sum of $A$ proton showers of energy $E=E_0/A$.  The
shower maximum $t$ penetrates into the atmosphere as $t \propto\ln E$, hence,
most air shower observables at ground level scale proportional to $\ln A$.
Several experiments publish observables, from which the mean logarithmic mass
can be derived.

\begin{table*}[hbt]
 \caption{\small Experiments and their corresponding air shower components analyzed 
          to derive the mean logarithmic mass: 
	  e: electromagnetic, $\mu$: muonic, h: hadronic
	  component, \v{C}: \v{C}erenkov, and F: fluorescence light. 
	  The energy threshold for muons and hadrons is given.
	  The last column lists the model used to interpret the data.} 
 \label{massetab}  
 \begin{center}
 \renewcommand{\arraystretch}{1.1}

 \begin{tabular}{lccccl}\hline
  Experiment                   &e&$\mu~(E_{th})$&\v{C}&F&
       model to derive $X_{max}$  \\ \hline
  BLANCA \cite{blanca}         &x&            &x& &CORSIKA QGSJET \\
  CACTI  \cite{cacti}          &x&            &x& &MOCCA--92      \\
  DICE \cite{dice}             &x& 800~MeV    &x& &CORSIKA VENUS  \\
  Fly's Eye \cite{flyseye}     & &            & &x&KNP \\
  Haverah Park \cite{haverah}  &x&     & & &AIRES/SIBYLL\\
  HEGRA (Airobic) \cite{hegra} &x&            &x& &CORSIKA QGSJET \\
  HiRes/MIA \cite{hiresmasse}  & & 800~MeV    & &x&CORSIKA QGSJET \\
  Mt. Lian Wang \cite{lianwang}&x&        &x& &MC with minijet model\\
  SPASE/VULCAN \cite{spase}    &x&            &x& &MOCCA SIBYLL\\
  Yakutsk \cite{yakumasse}     &x&            &x& &QGSJET\\ \hline
 \end{tabular}\vspace*{\baselineskip}

 \begin{tabular}{lcccl}\hline
  Experiment & e &$\mu~(E_{th})$&h~$(E_{th})$& air shower model \\\hline
  CASA--MIA \cite{casam} & x & 800~MeV    &       &MOCCA $\rightarrow$ QGSJET\\
  Chacaltaya \cite{chacaltaya}& x &            & 5~TeV       &CORSIKA QGSJET\\
  EAS--TOP/MACRO \cite{eastopmacro}& x &1.3~TeV &            &CORSIKA QGSJET\\
  EAS--TOP   \cite{eastopem} & x & 1~GeV      &              &CORSIKA QGSJET \\
  HEGRA (CRT) \cite{crt}     & x & track angle &             &CORSIKA VENUS\\
  KASCADE (hadrons/muons) \cite{jrhmasse}& &230~MeV  & 50~GeV&CORSIKA QGSJET\\
  KASCADE (electrons/muons) \cite{ulrich}&x&230~MeV  &       &CORSIKA QGSJET\\
  KASCADE (neural network) \cite{roth}&x&230/2400~MeV&100~GeV&CORSIKA QGSJET\\
  \hline
 \end{tabular} 

 \end{center}
\end{table*}

\begin{figure*}[hbt]\centering
 \epsfig{file=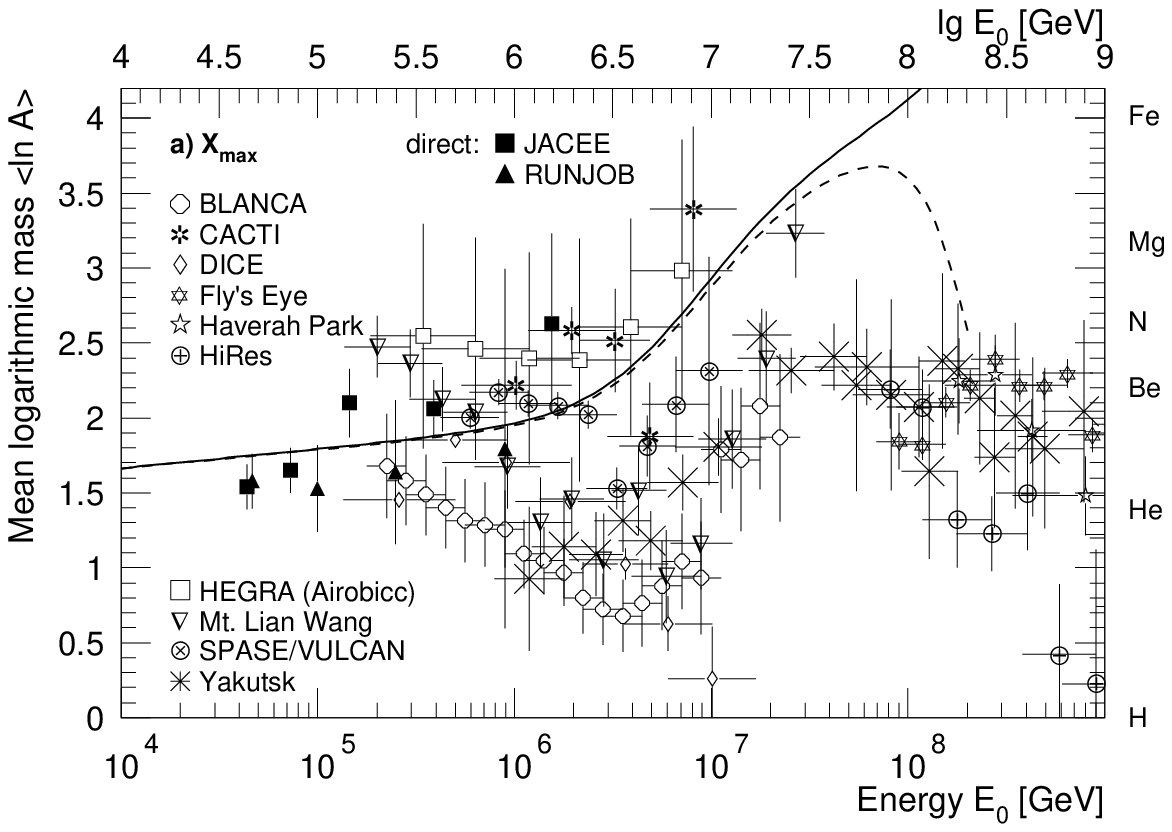,width=12cm}
 \epsfig{file=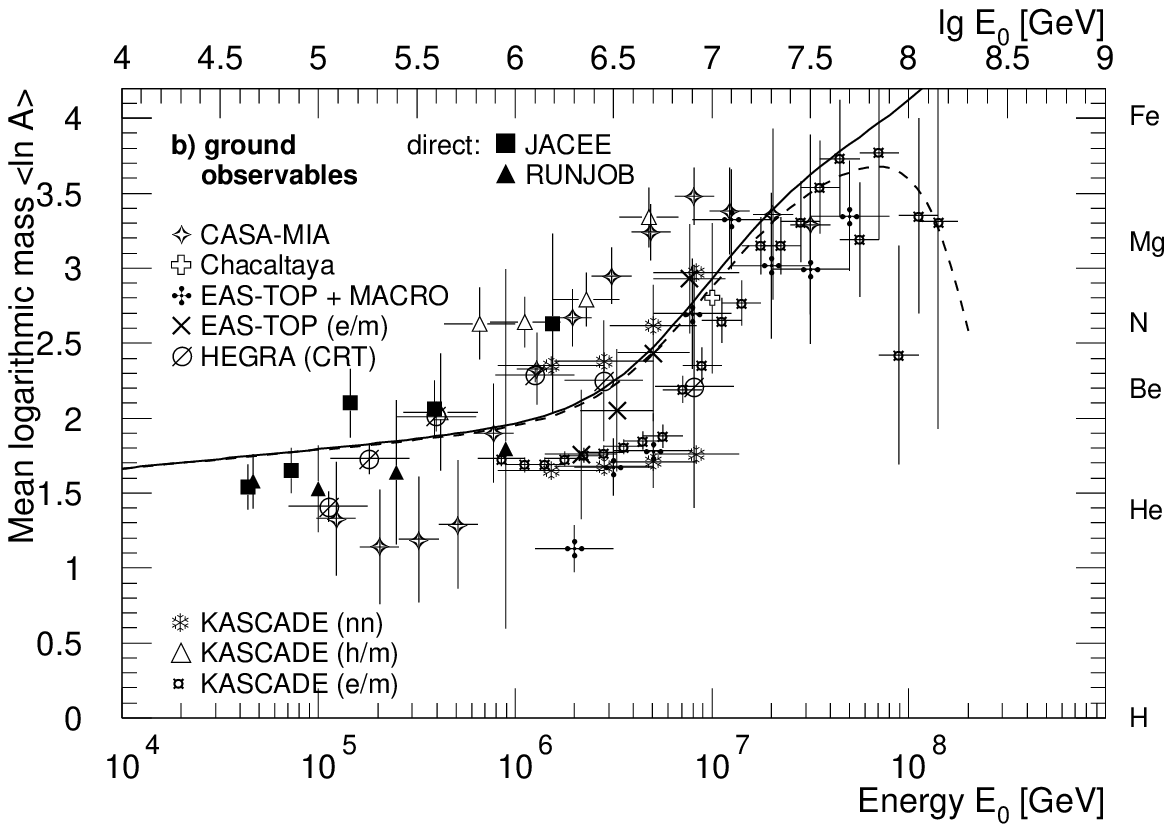,width=12cm}
 \caption{\small Mean logarithmic mass vs. primary energy.
	  a) Results from the average depth of the shower maximum $X_{max}$
	  using CORSIKA/QGSJET simulations.  b) Results from measurements of
	  distributions for electrons, muons, and hadrons at ground level.  For
	  references see Table~\ref{massetab}.  Results from the balloon
	  experiments JACEE \cite{jaceemasse} and RUNJOB \cite{runjoball} are
	  given as well.  Predictions according to the \modell\ are represented
	  by the solid line. The dashed lines are obtained by introducing an
	  {\sl ad--hoc} component of hydrogen only, see text.  }
 \label{masse}
\end{figure*}

\subsection{Average depth of shower maximum}
A typical observable is the average depth of the shower maximum $X_{max}$ in
the atmosphere.  Figure~\ref{xmax} summarizes recent measurements using
\v{C}erenkov and fluorescence light observations. In addition, results from
Haverah Park are presented.  This experiment measures the electromagnetic
component only, no \v{C}erenkov or fluorescence light.  $X_{max}$ is derived
from the structure of the arrival time distribution and the lateral
distribution of secondary particles.  Characteristics of the experiments are
compiled in the first part of Table~\ref{massetab}.

The results show systematic differences of $\approx 30$~g/cm$^2$ at 1~PeV
increasing to $\approx 65$~g/cm$^2$ close to 10~PeV.  The strongest increase of
$X_{max}$ as function of energy is reported by the DICE experiment.  A recent
reanalysis indicates, that part of this behaviour could be caused by systematic
effects, including electronic saturation in the photomultiplier tubes, which
had not been taken into account in the original analysis \cite{dicereanalyse}.

Only the Fly's Eye, HiRes, and DICE experiments observe an image of the air
shower and derive the average depth of the shower maximum directly.  All other
experiments use non--imaging techniques, i.e. the average depth of the shower
maximum is derived from properties of secondary particle distributions at
ground.  Hence, the results obtained depend on the models used to describe the
air shower development.  The programs applied by different groups are listed in
Table~\ref{massetab}.  Many groups use CORSIKA with the interaction models
QGSJET or VENUS, but AIRES, KNP, and MOCCA are also applied.  A private model
including minijets is used for the Mt.~Lian Wang analysis.  Application of
different codes can explain parts of the discrepancies in $X_{max}$.  For
example, Dickinson et al. found that using different interaction models in the
MOCCA code, viz. the original and the SIBYLL model, the systematic error
amounts to $\Delta X_{max}\approx 10$~g/cm$^2$ \cite{spase}.

The measurements are compared to predictions of the shower maximum for protons
and iron nuclei from simulations by Heck et al., using CORSIKA/QGSJET
\cite{heck}, and by Pryke, using the MOCCA code with the internal hadronic
interaction model \cite{pryke}.  In the MOCCA model the energy is transported
deeper into the atmosphere and the shower maxima are shifted by about
50~g/cm$^2$ to larger $X_{max}$ values as compared to CORSIKA/QGSJET.  This
results in a heavier composition. If the mean logarithmic mass is derived from
$X_{max}$ according to the MOCCA simulations, the data above 10~PeV indicate a
pure Fe composition.  As mentioned above CORSIKA/QGSJET seems to describe the
development of air showers best.  Hence, in the following the QGSJET
interpretation is used to derive $\lna$.

Despite of the differences mentioned, a general trend is visible in the
data. All results start with an intermediate composition. However, the measured 
$X_{max}$ increase faster as function of energy than the model predictions,
i.e. the measured values approach the proton distribution up to about 4~PeV.
Above this energy, a change in the shape of the measured $X_{max}$ curves
can be recognized in Figure~\ref{xmax} and the data start to approximate the
calculations for iron nuclei with a nearest approach at about 30~PeV. 
Finally, at higher energies the measurements move again in
direction of the proton predictions.

Knowing the average depth of the shower maximum for protons $X_{max}^p$ and 
iron nuclei $X_{max}^{Fe}$ from simulations,
the mean logarithmic mass can be derived in the superposition model of 
air showers from the measured $X_{max}^{meas}$ using 
\begin{equation}\label{xmaxfun}
\lna=\frac{X_{max}^{meas}-X_{max}^{p}}{X_{max}^{Fe}-X_{max}^{p}} 
   \cdot \ln A_{Fe}\quad .
\end{equation}
The corresponding $\lna$ values, obtained from the results shown in
Figure~\ref{xmax}, are plotted in Figure~\ref{masse}\,a  versus the primary
energy.  One observes a large scattering of the values. In the region 1 to
10~PeV the maximum variations amount to $\Delta\lna\approx2$.

\subsection{Particle distributions at ground level}
Results on $\lna$ from experiments measuring the electromagnetic, muonic, and
hadronic components at ground level are shown in Figure~\ref{masse}\,b.  The
experiments and the observables used as well as the air shower model applied
are listed in the second part of Table~\ref{massetab}.  The number of electrons
and the number of muons are used by CASA--MIA, EAS--TOP/MACRO, and KASCADE
(electrons/muons).  EAS--TOP uses the number of muons at a distance of 200~m to
the shower core in addition to the number of electrons.  The hadronic component
is analyzed by the Chacaltaya and KASCADE groups.  HEGRA (CRT) measures the
longitudinal shower development using the average production height of muons,
i.e. the angle of incidence for individual muon tracks.  Most experiments use
parametric methods but also classifications on an event--by--event--basis are
applied, e.g. in the KASCADE neural network analysis.

Several groups study the systematic effect caused by different interaction
models.  If possible, the CORSIKA QGSJET interpretation is utilized for the
compilation.  CASA--MIA uses MOCCA/SIBYLL simulations but in their analysis the
results have been modified in order to match results from QGSJET \cite{casam}.

In the KASCADE neural network analysis \cite{roth} several combinations of
observables and their influence on $\lna$ have been investigated.  The
observables include the number of electrons,  the number of muons for energy
thresholds of 230~MeV and 2.4~GeV, the number of hadrons with $E_h>100$~GeV and
their energy sum as well the spatial structure of high--energy muons.  In the
figure two extreme cases are shown with a difference of about 0.7 to 1.5 in
$\lna$. The lower values are obtained by analyzing the number of electrons and
the number of high--energy muons, while the high values result from
investigations of the number of low--energy muons, the hadronic energy sum, and
the spatial structure of the high--energy muons.

In another investigation of several hadronic observables by the KASCADE group
the systematic differences amount to $\Delta\ln A\approx 0.4$ \cite{jrhmasse},
compatible with an estimate of Swordy et al. claiming an uncertainty of
$\sigma\lna\approx 0.23$ \cite{kneeworkshop}.  This, most likely, represents
the systematic uncertainties introduced by inconsistencies in the hadronic
interaction models and M.C. programs used to interpret the data.

In both graphs of Figure~\ref{masse} the mean logarithmic mass calculated
according to the \modell\ with rigidity dependent cut--off is given by the
solid line.  Since more and more elements reach their cut--off energy
$\hat{E}_Z$, the mean logarithmic mass increases with rising energy and would
finally reach pure uranium.

As can be seen in Figure~\ref{flyseye1} and has been discussed in
section~\ref{edependence}, above 100~PeV the sum spectrum for $1\le Z\le92$
does not sufficiently describe the all--particle spectrum.  For this reason, a
new component is introduced {\sl ad hoc} in order to fill the difference
between the sum spectrum (solid graph) and the all--particle spectrum
represented by the dashed line in Figure~\ref{flyseye1}. Several theories
predict an extragalactic component consisting of hydrogen and helium nuclei
only, see e.g. \cite{biermann}.  
In fact, the AKENO experiment \cite{hara} found that the mass composition
becomes heavier with energy up to values of $\lna\approx3$ at
$N_e\approx10^{7.5}$.  For larger showers the composition gets lighter again,
and in analyses with two or three mass groups the proton component recovers.
The conversion of $N_e$ to primary energy had not been done, but cautious
estimations relate $\lg N_e=7.5$ to about 300~PeV.  

Anticipating a pure proton composition yields a lower limit for the $\lna$
values independent of any model assumptions.  Hence, only protons are assumed
for the {\sl ad--hoc} component.  The mean logarithmic mass obtained is
presented in Figure~\ref{masse} as dashed lines.

The results from experiments measuring electrons, muons and hadrons at ground
level are in reasonable agreement with direct measurements in the region of
overlap, namely below 1~PeV. Up to about 3~PeV they show a small trend of an
increasing $\lna$. For higher energies the measured $\lna$ values indicate a
more pronounced increase. Around 30~PeV values of $\lna\approx3.3$,
corresponding to silicon, are obtained.  The prediction of the \modell\ yields
the same shape of $\lna$ as function of energy.  On the other hand, systematic
differences between the results derived from the $X_{max}$ observations and the
model predictions are apparent.

\subsection{$X_{max}$ versus ground level observables} \label{xmaxvsground}
\begin{figure}[hbt]
 \epsfig{file=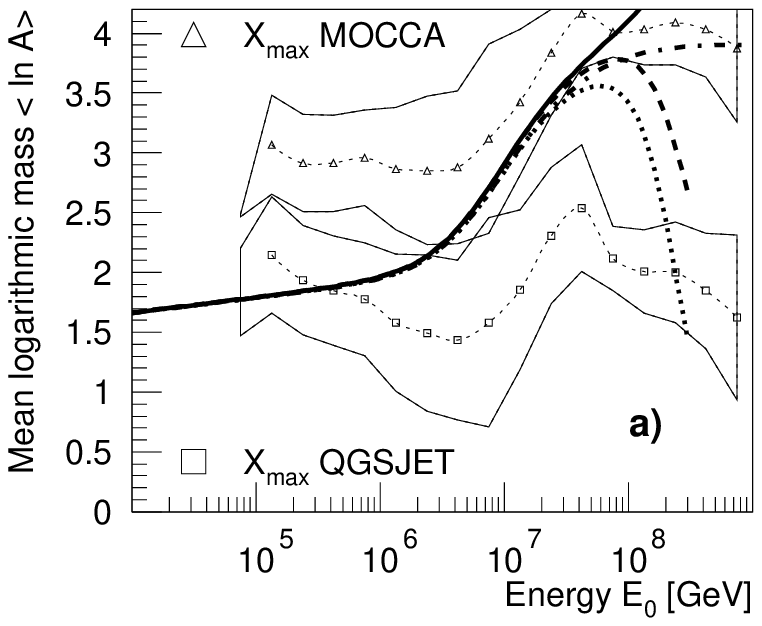,width=\columnwidth}
 \epsfig{file=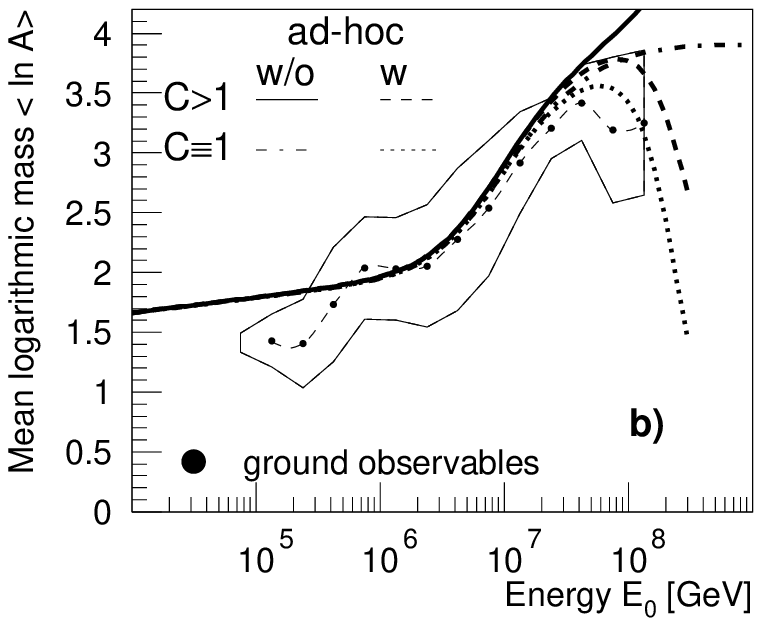,width=\columnwidth}
 \caption{\small Average mean logarithmic mass $\lna$ vs. primary energy.
	  Predictions of the \modell\ are given without and with the {\sl
	  ad-hoc} component as well as for the non--linear ($C>1$) and linear
	  ($C\equiv1$) extrapolation of $\gamma_Z$.  They are compared to
	  averaged values for experiments: a) observing $X_{max}$ using two
	  different interaction models for the interpretation,  b) measuring
	  electrons, muons and hadrons at ground level.  The contour lines
	  represent the r.m.s. values of the data.}
 \label{xmaxlna0}
\end{figure}

In order to study the difference in shape and absolute value of $\lna$ in more
detail, the weighted average of $\lna$ is calculated for both, ground
observables and $X_{max}$ measurements, taking into account the given errors
for the individual measurements.  For the calculation of $\lna$ from $X_{max}$
simulations with CORSIKA/QGSJET \cite{heck} and the MOCCA code \cite{pryke} are
used.  Both results are shown in Figure~\ref{xmaxlna0}\,a together with the
corresponding average values for ground observables in
Figure~\ref{xmaxlna0}\,b.  The contour lines represent the r.m.s. values of
the data.

The parameters of the \modell\ have been obtained by fits to the all--particle
energy spectrum. The mean logarithmic mass is an independent observable, which
allows an additional check of the model. The values predicted by the model are
compared to the data in Figure~\ref{xmaxlna0}. To estimate the systematic
uncertainties introduced by the extrapolation of $\gamma_Z$ for heavy nuclei,
the mean logarithmic mass is shown for the non--linear ($C>1$) and the linear
($C\equiv1$) extrapolation by the solid and dash dotted lines, respectively.
In addition, $\lna$ values including an {\sl ad--hoc} component of protons only
are represented for both cases by the dashed and dotted lines, respectively.
The differences between the two extrapolations for $\gamma_Z$ are relative
small, e.g. the maximum $\lna$ values for the two cases including the {\sl
ad--hoc} component differ only by $\Delta\lna\approx0.2$.

The variations of $\lna$ derived from ground observables are in the same order
as the uncertainties due to the observables used. The investigations by the
KASCADE group --- as discussed above --- suggest a systematic error of about
0.5 to 0.7 for the determination of $\lna$ compatible with the average
variation of $\approx0.5$ in Figure~\ref{xmaxlna0}\,b. The results of the
\modell\ are in reasonable agreement with the measurements, absolute values and
the shape are almost the same.  Even the maximum of $\lna$ around 70~PeV is
reflected by the measurements. Below 0.5~PeV only a few indirect measurements
contribute, see Figure~\ref{masse}, and the deviations between the average data
and the \modell\ are not to be taken seriously. Above 20~PeV the model yields a
heavier composition as compared to the mean value of the measurements.  It can
not be excluded, that in some analysis procedures the mass values had been
restricted to a range limited by the extreme values pure protons and pure iron
nuclei. Therefore, the numbers at high energies could be biased towards a
lighter composition.

The mean logarithmic mass derived from shower maximum observations shows for
both simulation codes below the {\sl knee} energy a slightly falling $\lna$.
Beyond this energy $\lna$ increases, but less pronounced as compared to the
ground observations. The $\lna$ values reach their maximum at about 50~PeV.
This energy is in agreement with the phenomenological model, where the maximum
is caused by the introduction of the {\sl ad--hoc} component.  As can be
inferred from Figure~\ref{xmaxlna0}\,a, the results of the $X_{max}$
measurements are not compatible with the \modell. Especially below the {\sl
knee} they do not match the results of direct observations, nor their
extrapolations to higher energies. The decreasing $\lna$ values below the {\sl
knee} are not seen by direct measurements.  This suggests a systematic
difference between ground observables and $X_{max}$ measurements which could
have its origin in the air shower simulations or data analysis procedures
applied.  The effect is independent of the simulation program used.  Both,
CORSIKA and MOCCA, yield different absolute values, but the same inclined "S"
shape is obtained.

\subsection{Relation for cut--off energy}
\begin{figure}[hbt]
 \epsfig{file=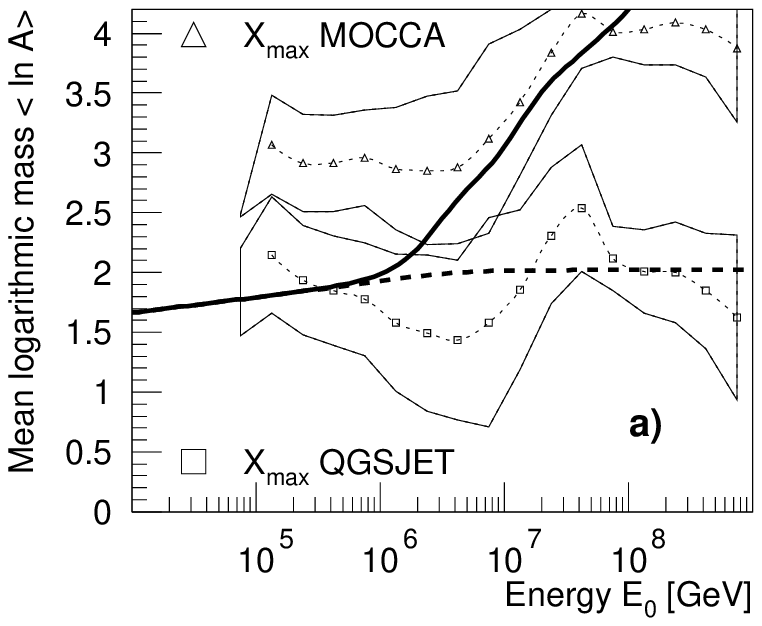,width=\columnwidth}
 \epsfig{file=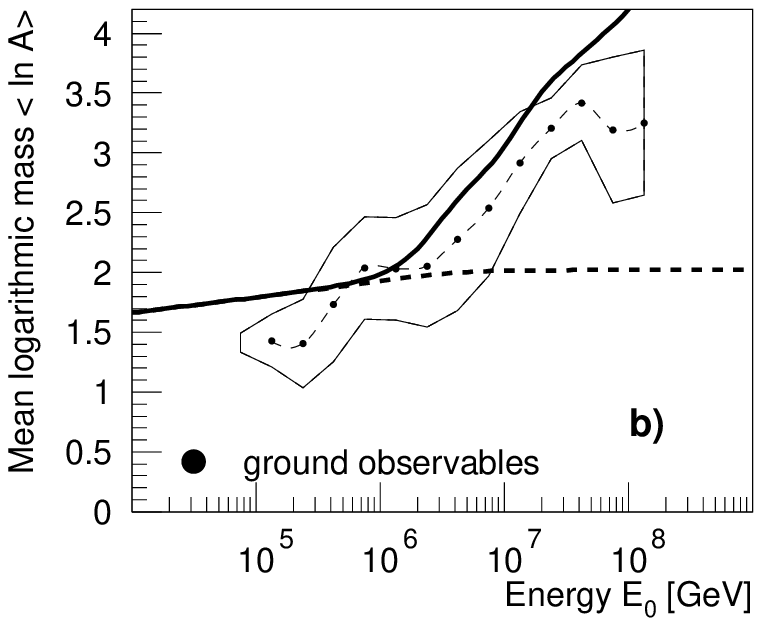,width=\columnwidth}
 \caption{\small Average mean logarithmic mass $\lna$ vs. primary energy.
	  Predictions of the \modell\ with mass dependent cut--off (solid line)
	  and with constant cut--off (dashed line) are compared to averaged
	  experimental values. The experimental data in a) and b) are the same
	  as in Figure~\ref{xmaxlna0}.}
 \label{xmaxlna1}
\end{figure}

The mean logarithmic mass calculated according to the \modell\ with mass
dependent cut--off shows a similar shape but yields larger values as compared
to a rigidity dependent cut-off.  The results for a common $\gamma_c$ using the
non--linear extrapolation for $\gamma_Z$ are plotted in Figure~\ref{xmaxlna1}
versus the particle energy and compared to the average results of the
experiments.  While the $\lna$ values for the rigidity dependent cut--off are
very close to the average measured results, see Figure~\ref{xmaxlna0}\,b, the
values for a mass dependent cut--off above 1~PeV are about 0.3 larger than the
average data.  Thus, a rigidity dependent cut--off is favoured against the mass
dependent approach.  The hypothesis of a constant cut--off energy yields an
almost constant mean logarithmic mass, as shown in Figure~\ref{xmaxlna1}, and
is not compatible with the measurements. 

Combining both, the arguments discussed above in connection with the shape of
the all--particle energy spectrum and the findings concerning the mean
logarithmic mass, the \modell\ with rigidity dependent cut--off, the
non--linear extrapolation of $\gamma_Z$, and a common $\Delta\gamma$ is the
favoured solution.  The $\lna$ values versus energy for this model are
summarized in Table~\ref{mtab} for reference.

\section{Compatibility with results from indirect measurements}
\label{compat}
The parameters of the \modell\ have been determined by extrapolating the
directly measured energy spectra to high energies and a fit to the
all--particle spectrum of indirect mesurements in section \ref{espec}. As a
first independent test, the mean logarithmic mass calculated with the \modell\
has been compared to experimental results in section \ref{secmasse}. In the
following, additional checks of consistency are presented.  Two results from
air shower measurements are compared to the predictions of the model.  Namely
the spectra for mass groups obtained by the KASCADE experiment \cite{ulrich}
and $X_{max}$ distributions published by the Fly's Eye group \cite{gaisser}
to test the predictions in the energy regions of the {\sl knee} and the
cut--off for the ultra--heavy elements, respectively.

\subsection{Energy spectra for groups of elements}
\begin{figure}[hbt]
 \epsfig{file=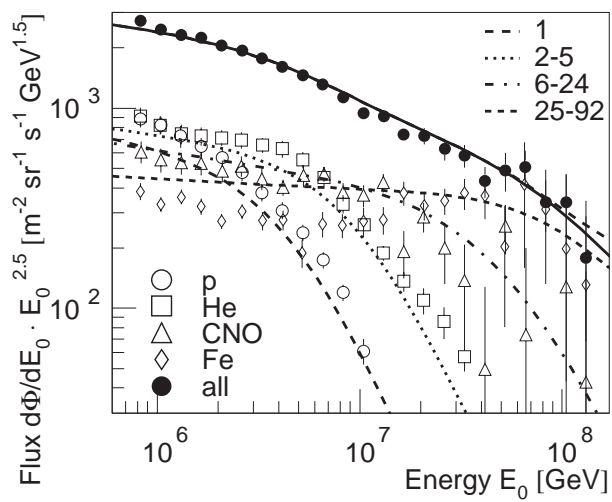,width=\columnwidth}
 \caption{\small Energy spectra for four elemental groups and the all--particle
   spectrum as measured by the KASCADE experiment \cite{ulrich} (symbols)
   compared to the spectra of the \modell\ (lines).}
 \label{ulrich}
\end{figure}

Recently, individual spectra for groups of elements in the energy range from
$10^6$~GeV to $10^8$~GeV have been reconstructed from air shower measurements
by the KASCADE group \cite{ulrich}.  Using shower size spectra for the
electromagnetic and muonic component for three different zenith angle bins the
primary energy spectra for elemental groups have been obtained by an unfolding
procedure. Four groups, represented by hydrogen, helium, CNO, and iron have
been chosen.  These groups are identified with the following groups in the
\modell: $Z=1$, $2\le Z\le5$, $6\le Z\le24$, and $25\le Z\le92$.  Renormalizing
the KASCADE energy scale (according to Table~\ref{eshift} by -7\%) the spectra
shown in Figure~\ref{ulrich} are obtained.

The measured all--particle spectrum agrees well with the calculated flux also
shown in the figure.  The shapes of the experimental spectra for protons and
the helium group are compatible with the spectra calculated according to the
\modell. It is remarkable that the cut--off behaviour of both components agrees
well with the calculated spectra, thus independently confirming the cut--off
parameters obtained in section~\ref{espec}. The measured spectra for the CNO
group and the iron group also agree with the calculations, with somewhat larger
experimental error bars at high energies. In the energy region around
$10^6$~GeV the KASCADE fluxes for protons and helium are slightly larger than
in the \modell\ and the values for the iron group are somewhat below the
calculated flux.  On the whole, in an overall view the spectra of the \modell\
agree well with the unfolding analysis.
It may be pointed out that both analyses are completely independent and they
yield very similar results.

\subsection{$X_{max}$--distributions}
The energy region between $10^8$~GeV and $10^9$~GeV where the heaviest elements
reach their cut--off energies and the {\sl ad--hoc} component starts to
dominate the all--particle spectrum is an interesting range for further
independent tests of the predictions of the model.  

Assuming protons only for the {\sl ad--hoc} component, relative abundances for
elemental groups are obtained as compiled in Table~\ref{fracttab} for energies
$E_0=3\cdot10^8$~GeV, $5\cdot10^8$~GeV, and $10^9$~GeV.  In the energy range
mentioned the fraction of heavy and ultra--heavy elements is expected to
decrease from 62\% to 35\%. On the other hand the fraction of the {\sl ad--hoc}
component steadily increases, accounting for the rise of the light elements
(protons).

\begin{table}[hbt]
 \caption{\small Relative abundances for groups of nuclei with charge number $Z$ for
 different primary energies $E_0$ according to the \modell. The {\sl ad--hoc}
 component is assumed to be protons only.}
 \label{fracttab}

 \renewcommand{\arraystretch}{1.1}
 \begin{tabular}{lcrrr} \hline
 &&\multicolumn{3}{c}{Energy $E_0$/GeV}\\ \cline{3-5}
 & $Z$ & $3\cdot10^8$ & $5\cdot10^8$ & $10^9$ \\ \hline
 light & 1-12 & 38\% & 52\% & 65\% \\
 heavy &13-29 & 24\% & 12\% &  8\% \\
 ultra heavy & 30-92 & 38\% & 36\% & 27\% \\
 \hline
 \end{tabular}
\end{table}

\begin{figure}[hbt]
 \epsfig{file=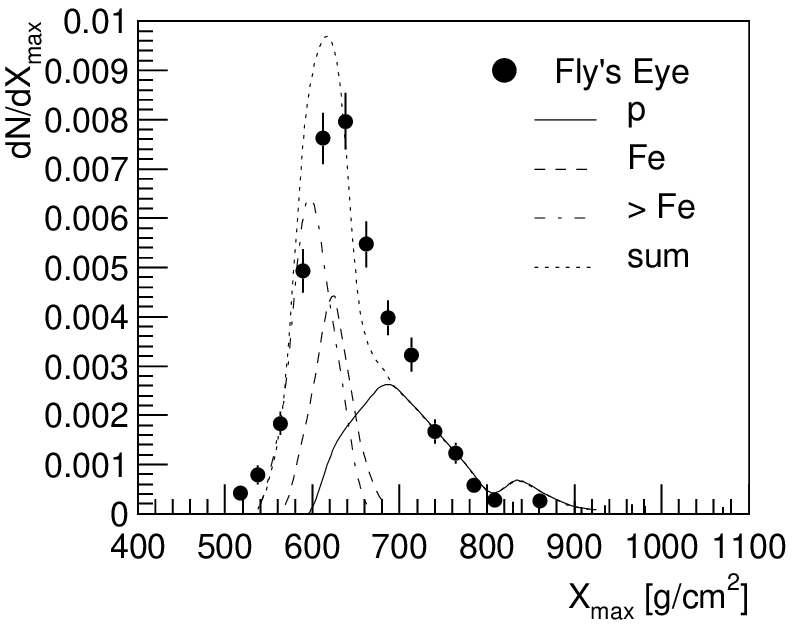,width=0.92\columnwidth}
 \epsfig{file=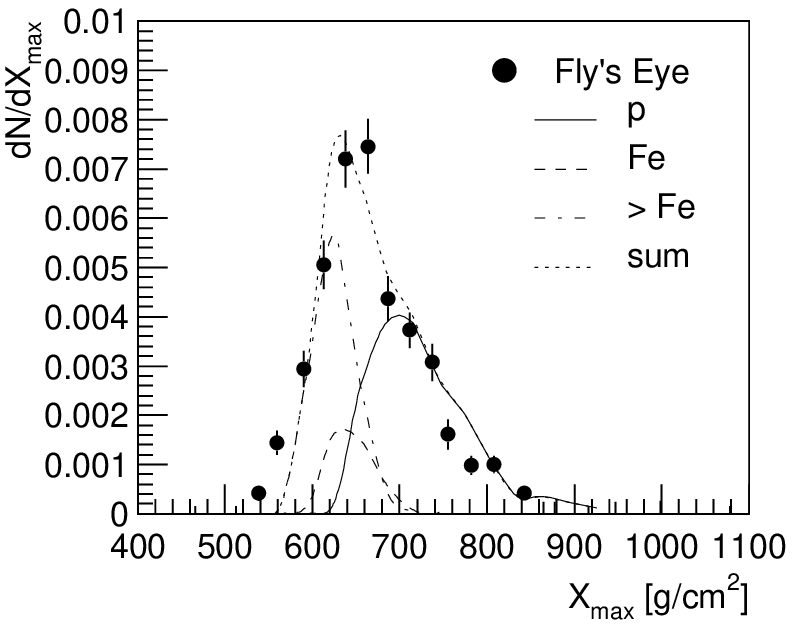,width=0.92\columnwidth}
 \epsfig{file=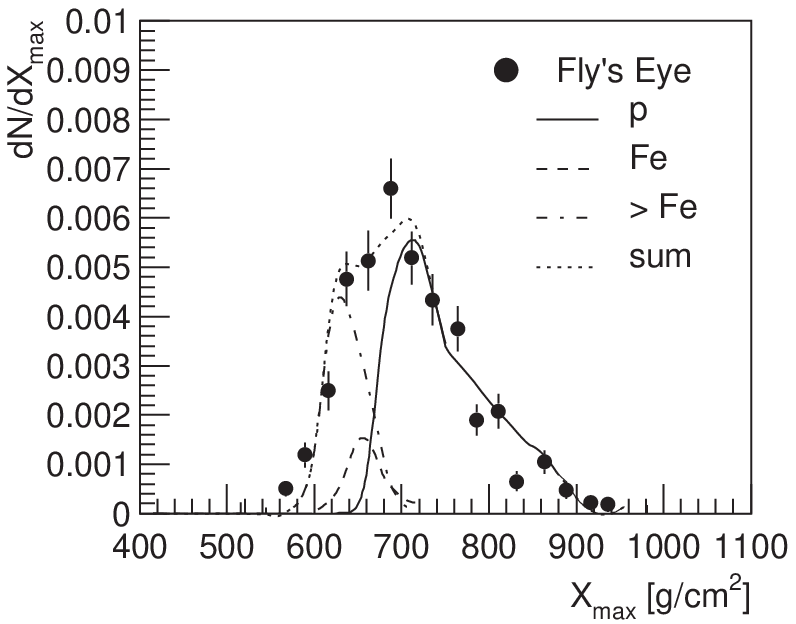,width=0.92\columnwidth}
 \caption{\small Distributions for the average depth of the shower maximum $X_{max}$
   for different energy ranges: 3 to $5\cdot10^8$~GeV (top), $5\cdot10^8$~GeV
   to $10^9$~GeV (middle), and above $10^9$~GeV (bottom).  The measured values
   of the Fly's Eye experiment \cite{gaisser} are compared to simulated values
   for protons, iron and ultra--heavy nuclei, see text.}
 \label{gaisser}
\end{figure}

Adopting the mass composition listed in Table~\ref{fracttab}, simulations have
peen performed in order to study the expected distributions of the average
depth of the shower maximum $X_{max}$.  The air shower simulation program
CORSIKA (6.0120) with the interaction model QGSJET has been used to calculate
$X_{max}$ for primary protons and iron induced showers for the energies listed
in the table. For each energy and species 150 showers have been generated. The
CORSIKA code allows primary nuclei with mass numbers $A\le56$. The $X_{max}$
distributions for ultra--heavy nuclei have been estimated using the assumption,
$X_{max}\propto\ln A$, based on the simulated values for protons and iron
nuclei.

In section~\ref{xmaxvsground} it has been discussed, that there are
discrepancies in the mean logarithmic mass calculated from electron, muon, and
hadron distributions on one hand and $X_{max}$ values on the other hand.
Therefore, it can not be expected, that the CORSIKA predictions match the
measured values.  The distances between the lines in Figure~\ref{xmax} suggest,
that the differences between the average $X_{max}$ values for protons and iron
nuclei are almost independent from the interaction model.  If a constant offset
is added to the simulated values, only the absolute values of the average
depths of the shower maximum are shifted but the fluctuations stay the same.
Hence, to reconcile the simulations with the measured data, an offset $\Delta
X_{max}=30$~g/cm$^2$ has been added to the simulated values.  The resulting
distributions are compared in Figure~\ref{gaisser} to measurements of the Fly's
Eye group \cite{gaisser}.  

As can be inferred from the figure, with this offset the simulated and measured
distributions are compatible with each other.  For the lowest energy bin the
assumption of two groups only for elements up to iron is only a crude
approximation of the real mass composition and the simulation of intermediate
elements like the CNO group would improve the agreement between the simulated
and measured distributions.  The structures on the right hand side of the
proton distributions are due to fluctuations in the simulated $X_{max}$ values.

The measured showers seem to penetrate deeper into the atmosphere than
predicted by the CORSIKA calculations but the width of the measured
distributions, i.e. the $X_{max}$ fluctuations are well compatible with the
mass composition predicted by the \modell.

\section{Conclusion}
An all--particle primary energy spectrum has been combined from many air shower
experiments by properly adjusting the individual energy scales to meet the flux
values of direct measurements.  The resulting power law spectrum exhibits two
changes in the spectral slope at 4.5~PeV and about 0.4~EeV.  

A phenomenological model, named \modell, is adopted in which the energy
spectra of individual nuclei have a power law behaviour with a cut--off at a
specific energy $\hat{E}_Z$. The {\sl knee} is explained as the subsequent
cut--offs of the individual elements of the galactic component, starting with
protons. The second {\sl knee} seems to indicate the end of the stable elements
of the galactic component.  

It turns out, that the contribution of ultra--heavy elements with $Z>28$
is not negligible in the 100~PeV regime.
Their flux values have been calculated, taking into consideration 
the modulation in the heliosphere and power law spectra, for which the
spectral indices have been estimated in two independent ways.

Several assumptions about the cut--off behaviour for the individual elements
have been checked against the experimental data. It turns out that the
hypothesis the {\sl knee} energy scales with the charge of the individual
nuclei, $\hat{E}_Z=Z\cdot\hat{E}_p$, describes the data best.

Comparing several methods of extracting a mean logarithmic mass from the data
indicates a systematic error of $\Delta\lna\approx0.5$ for the determination of
$\lna$.  Taking this error into account, the mass composition calculated with
the \modell\ is in agreement with results from air shower experiments measuring
the electromagnetic, muonic and hadronic components at ground level.  But the
mass composition disagrees with results from experiments measuring the average
depth of the shower maximum with \v{C}erenkov and fluorescence detectors.

\section*{Acknowledgment}
The author would like to thank J.~Engler, K.--H.~Kampert, and J.~Knapp for 
critically reading the manuscript and many useful discussions
as well as the anonymous referee for helpful suggestions.

\begin{appendix}
\section{Tables}

\begin{table*}[p]
 \caption{\small Absolute flux $\Phi_Z^0~[(\mbox{m}^2\mbox{~sr~s~TeV})^{-1}]$ 
 at $E_0=1$~TeV/nucleus and spectral index $\gamma_Z$ of cosmic--ray elements.}
 \label{directtab}
 \renewcommand{\arraystretch}{1.1} 
 \begin{tabular}{rlcc c rlcc c rlcc}
  \hline  
   $Z$ &  & $\Phi_Z^0$ &$-\gamma_Z$ &&
   $Z$ &  & $\Phi_Z^0$ &$-\gamma_Z$ &&
   $Z$ &  & $\Phi_Z^0$ &$-\gamma_Z$ \\
  \cline{1-4}\cline{6-9}\cline{11-14}
  1\footnotemark[2]& H&$8.73\cdot 10^{-2}$&2.71 && 32\footnotemark[4]&Ge&$4.02\cdot 10^{-6}$&2.54 && 63\footnotemark[4]&Eu&$1.58\cdot 10^{-7}$&2.27\\ 
  2\footnotemark[2]&He&$5.71\cdot 10^{-2}$&2.64 && 33\footnotemark[4]&As&$9.99\cdot 10^{-7}$&2.54 && 64\footnotemark[4]&Gd&$6.99\cdot 10^{-7}$&2.25\\ 
  3\footnotemark[3]&Li&$2.08\cdot 10^{-3}$&2.54 && 34\footnotemark[4]&Se&$2.11\cdot 10^{-6}$&2.53 && 65\footnotemark[4]&Tb&$1.48\cdot 10^{-7}$&2.24\\ 
  4\footnotemark[3]&Be&$4.74\cdot 10^{-4}$&2.75 && 35\footnotemark[4]&Br&$1.34\cdot 10^{-6}$&2.52 && 66\footnotemark[4]&Dy&$6.27\cdot 10^{-7}$&2.23\\ 
  5\footnotemark[3]& B&$8.95\cdot 10^{-4}$&2.95 && 36\footnotemark[4]&Kr&$1.30\cdot 10^{-6}$&2.51 && 67\footnotemark[4]&Ho&$8.36\cdot 10^{-8}$&2.22\\ 
  6\footnotemark[3]& C&$1.06\cdot 10^{-2}$&2.66 && 37\footnotemark[4]&Rb&$6.93\cdot 10^{-7}$&2.51 && 68\footnotemark[4]&Er&$3.52\cdot 10^{-7}$&2.21\\ 
  7\footnotemark[3]& N&$2.35\cdot 10^{-3}$&2.72 && 38\footnotemark[4]&Sr&$2.11\cdot 10^{-6}$&2.50 && 69\footnotemark[4]&Tm&$1.02\cdot 10^{-7}$&2.20\\ 
  8\footnotemark[3]& O&$1.57\cdot 10^{-2}$&2.68 && 39\footnotemark[4]& Y&$7.82\cdot 10^{-7}$&2.49 && 70\footnotemark[4]&Yb&$4.15\cdot 10^{-7}$&2.19\\ 
  9\footnotemark[3]& F&$3.28\cdot 10^{-4}$&2.69 && 40\footnotemark[4]&Zr&$8.42\cdot 10^{-7}$&2.48 && 71\footnotemark[4]&Lu&$1.72\cdot 10^{-7}$&2.18\\ 
 10\footnotemark[3]&Ne&$4.60\cdot 10^{-3}$&2.64 && 41\footnotemark[4]&Nb&$5.05\cdot 10^{-7}$&2.47 && 72\footnotemark[4]&Hf&$3.57\cdot 10^{-7}$&2.17\\ 
 11\footnotemark[3]&Na&$7.54\cdot 10^{-4}$&2.66 && 42\footnotemark[4]&Mo&$7.79\cdot 10^{-7}$&2.46 && 73\footnotemark[4]&Ta&$2.16\cdot 10^{-7}$&2.16\\ 
 12\footnotemark[3]&Mg&$8.01\cdot 10^{-3}$&2.64 && 43\footnotemark[4]&Tc&$6.98\cdot 10^{-8}$&2.46 && 74\footnotemark[4]& W&$4.16\cdot 10^{-7}$&2.15\\ 
 13\footnotemark[3]&Al&$1.15\cdot 10^{-3}$&2.66 && 44\footnotemark[4]&Ru&$3.01\cdot 10^{-7}$&2.45 && 75\footnotemark[4]&Re&$3.35\cdot 10^{-7}$&2.13\\ 
 14\footnotemark[3]&Si&$7.96\cdot 10^{-3}$&2.75 && 45\footnotemark[4]&Rh&$3.77\cdot 10^{-7}$&2.44 && 76\footnotemark[4]&Os&$6.42\cdot 10^{-7}$&2.12\\ 
 15\footnotemark[3]& P&$2.70\cdot 10^{-4}$&2.69 && 46\footnotemark[4]&Pd&$5.10\cdot 10^{-7}$&2.43 && 77\footnotemark[4]&Ir&$6.63\cdot 10^{-7}$&2.11\\ 
 16\footnotemark[3]& S&$2.29\cdot 10^{-3}$&2.55 && 47\footnotemark[4]&Ag&$4.54\cdot 10^{-7}$&2.42 && 78\footnotemark[4]&Pt&$1.03\cdot 10^{-6}$&2.10\\ 
 17\footnotemark[3]&Cl&$2.94\cdot 10^{-4}$&2.68 && 48\footnotemark[4]&Cd&$6.30\cdot 10^{-7}$&2.41 && 79\footnotemark[4]&Au&$7.70\cdot 10^{-7}$&2.09\\ 
 18\footnotemark[3]&Ar&$8.36\cdot 10^{-4}$&2.64 && 49\footnotemark[4]&In&$1.61\cdot 10^{-7}$&2.40 && 80\footnotemark[4]&Hg&$7.43\cdot 10^{-7}$&2.08\\ 
 19\footnotemark[3]& K&$5.36\cdot 10^{-4}$&2.65 && 50\footnotemark[4]&Sn&$7.15\cdot 10^{-7}$&2.39 && 81\footnotemark[4]&Ti&$4.28\cdot 10^{-7}$&2.06\\ 
 20\footnotemark[3]&Ca&$1.47\cdot 10^{-3}$&2.70 && 51\footnotemark[4]&Sb&$2.03\cdot 10^{-7}$&2.38 && 82\footnotemark[4]&Pb&$8.06\cdot 10^{-7}$&2.05\\ 
 21\footnotemark[3]&Sc&$3.04\cdot 10^{-4}$&2.64 && 52\footnotemark[4]&Te&$9.10\cdot 10^{-7}$&2.37 && 83\footnotemark[4]&Bi&$3.25\cdot 10^{-7}$&2.04\\ 
 22\footnotemark[3]&Ti&$1.14\cdot 10^{-3}$&2.61 && 53\footnotemark[4]& I&$1.34\cdot 10^{-7}$&2.37 && 84\footnotemark[4]&Po&$3.99\cdot 10^{-7}$&2.03\\ 
 23\footnotemark[3]& V&$6.31\cdot 10^{-4}$&2.63 && 54\footnotemark[4]&Xe&$5.74\cdot 10^{-7}$&2.36 && 85\footnotemark[4]&At&$4.08\cdot 10^{-8}$&2.02\\ 
 24\footnotemark[3]&Cr&$1.36\cdot 10^{-3}$&2.67 && 55\footnotemark[4]&Cs&$2.79\cdot 10^{-7}$&2.35 && 86\footnotemark[4]&Rn&$1.74\cdot 10^{-7}$&2.00\\ 
 25\footnotemark[3]&Mn&$1.35\cdot 10^{-3}$&2.46 && 56\footnotemark[4]&Ba&$1.23\cdot 10^{-6}$&2.34 && 87\footnotemark[4]&Fr&$1.78\cdot 10^{-8}$&1.99\\ 
 26\footnotemark[2]&Fe&$2.04\cdot 10^{-2}$&2.59 && 57\footnotemark[4]&La&$1.23\cdot 10^{-7}$&2.33 && 88\footnotemark[4]&Ra&$7.54\cdot 10^{-8}$&1.98\\ 
 27\footnotemark[3]&Co&$7.51\cdot 10^{-5}$&2.72 && 58\footnotemark[4]&Ce&$5.10\cdot 10^{-7}$&2.32 && 89\footnotemark[4]&Ac&$1.97\cdot 10^{-8}$&1.97\\ 
 28\footnotemark[3]&Ni&$9.96\cdot 10^{-4}$&2.51 && 59\footnotemark[4]&Pr&$9.52\cdot 10^{-8}$&2.31 && 90\footnotemark[4]&Th&$8.87\cdot 10^{-8}$&1.96\\ 
 29\footnotemark[4]&Cu&$2.18\cdot 10^{-5}$&2.57 && 60\footnotemark[4]&Nd&$4.05\cdot 10^{-7}$&2.30 && 91\footnotemark[4]&Pa&$1.71\cdot 10^{-8}$&1.94\\ 
 30\footnotemark[4]&Zn&$1.66\cdot 10^{-5}$&2.56 && 61\footnotemark[4]&Pm&$8.30\cdot 10^{-8}$&2.29 && 92\footnotemark[4]& U&$3.54\cdot 10^{-7}$&1.93\\ 
 31\footnotemark[4]&Ga&$2.75\cdot 10^{-6}$&2.55 && 62\footnotemark[4]&Sm&$3.68\cdot 10^{-7}$&2.28 && && & \\
  \hline
 \end{tabular}\\
 \footnote{a}{This work, see Figures~\ref{directP} to \ref{directFe} and text.}\\
 \footnote{b}{From Wiebel--Sooth et al. \cite{wiebel}.} \\
 \footnote{c}{This work, extrapolation for ultra--heavy elements, see text.}
\end{table*}

\begin{table*}[p]
  \caption{\small All--particle cosmic--ray energy spectrum.
  Normalized average flux from air shower experiments $\Phi_m$ 
  and flux $\Phi_c$ according to the \modell.
  The errors are r.m.s. values of the data for $\Phi_m$ and
  the difference between $\Phi_m$ and $\Phi_c$ for $\Phi_c$.
  ($\lg E_0$ [GeV] and
   $\Phi(E_0)\cdot E_0^{2.5}~[\mbox{GeV}^{1.5}/\mbox{m}^2~\mbox{sr s}]$.)}
  \label{etab} 
  \begin{center}
  \renewcommand{\arraystretch}{1.1}
  \begin{tabular}{rr@{$\pm$}rr@{$\pm$}rcrr@{$\pm$}rr@{$\pm$}r}
  \hline 
  $\lg E_0$ & \multicolumn{2}{c }{$\Phi_m(E_0)\cdot E_0^{2.5}$}  
            & \multicolumn{2}{c }{$\Phi_c(E_0)\cdot E_0^{2.5}$} &&
  $\lg E_0$ & \multicolumn{2}{c }{$\Phi_m(E_0)\cdot E_0^{2.5}$}  
            & \multicolumn{2}{c }{$\Phi_c(E_0)\cdot E_0^{2.5}$} \\
  \cline{1-5} \cline{7-11}
 4.0&\multicolumn{2}{c}{}&\multicolumn{2}{l}{4974} && 6.0& 2417& 207& 2395& 22\\
 4.1&\multicolumn{2}{c}{}&\multicolumn{2}{l}{4796} && 6.1& 2270& 212& 2292& 21\\
 4.2&\multicolumn{2}{c}{}&\multicolumn{2}{l}{4624} && 6.2& 2164& 177& 2182& 19\\
 4.3&\multicolumn{2}{c}{}&\multicolumn{2}{l}{4459} && 6.3& 2051& 191& 2065& 15\\
 4.4&\multicolumn{2}{c}{}&\multicolumn{2}{l}{4300} && 6.4& 1951& 203& 1939& 12\\
 4.5&\multicolumn{2}{c}{}&\multicolumn{2}{l}{4146} && 6.5& 1843& 230& 1802& 40\\
 4.6&\multicolumn{2}{c}{}&\multicolumn{2}{l}{3999} && 6.6& 1680& 236& 1658& 22\\
 4.7&  3911&   194 &  3857&   54 &&  6.7&  1519 &   245 &  1509 &   10  \\
 4.8&  3841&   201 &  3720&  121 &&  6.8&  1368 &   250 &  1362 &    7  \\
 4.9&  3711&   197 &  3589&  122 &&  6.9&  1229 &   213 &  1220 &    9  \\
 5.0&  3579&   194 &  3463&  116 &&  7.0&  1076 &   204 &  1090 &   14  \\
 5.1&  3427&   192 &  3341&   86 &&  7.1&  1011 &   229 &   974 &   38  \\
 5.2&  3345&   193 &  3224&  121 &&  7.2&   914 &   262 &   871 &   43  \\
 5.3&  3218&   191 &  3111&  108 &&  7.3&   815 &   333 &   782 &   33  \\
 5.4&  3086&   201 &  3002&   85 &&  7.4&   666 &   187 &   702 &   36  \\
 5.5&  2917&   251 &  2896&   21 &&  7.5&   586 &   115 &   629 &   43  \\
 5.6&  2772&   252 &  2794&   22 &&  7.6&   514 &    97 &   559 &   46  \\
 5.7&  2704&   237 &  2693&   11 &&  7.7&   461 &   100 &   493 &   33  \\
 5.8&  2645&   235 &  2594&   50 &&  7.8&   426 &    95 &   429 &    4  \\
 5.9&  2526&   217 &  2495&   31 &&  7.9&   353 &    75 &   369 &   16  \\
 \multicolumn{5}{c}{}            &&  8.0&   323 &    54 &   314 &    9  \\
 \hline 
  \end{tabular}
  \end{center}
\end{table*}

\begin{table*}[p]
  \caption{\small Mean logarithmic mass according to the \modell\ ($\lg E_0$ [GeV]).}
  \label{mtab} 
  \begin{center}
  \renewcommand{\arraystretch}{1.1}
  \begin{tabular}{rrcrrcrrcrr}
  \hline 
  $\lg E_0$ & $\lna$ && 
  $\lg E_0$ & $\lna$ && 
  $\lg E_0$ & $\lna$ && 
  $\lg E_0$ & $\lna$ \\
  \cline{1-2}\cline{4-5}\cline{7-8}\cline{10-11}
 4.0 &   1.66&& 4.1 &   1.68&& 4.2 &   1.69&& 4.3 &   1.70\\
 4.4 &   1.72&& 4.5 &   1.73&& 4.6 &   1.74&& 4.7 &   1.75\\
 4.8 &   1.77&& 4.9 &   1.78&& 5.0 &   1.79&& 5.1 &   1.81\\
 5.2 &   1.82&& 5.3 &   1.83&& 5.4 &   1.85&& 5.5 &   1.86\\
 5.6 &   1.88&& 5.7 &   1.90&& 5.8 &   1.92&& 5.9 &   1.94\\
 6.0 &   1.97&& 6.1 &   2.00&& 6.2 &   2.04&& 6.3 &   2.09\\
 6.4 &   2.16&& 6.5 &   2.24&& 6.6 &   2.35&& 6.7 &   2.47\\
 6.8 &   2.61&& 6.9 &   2.77&& 7.0 &   2.93&& 7.1 &   3.09\\
 7.2 &   3.24&& 7.3 &   3.38&& 7.4 &   3.51&& 7.5 &   3.62\\
 7.6 &   3.72\footnotemark[5]-3.73\footnotemark[6]&& 7.7 &   3.80-3.83&& 7.8 &   3.87-3.92&& 7.9 &   3.92-4.02\\
 8.0 &   3.93-4.12&& 8.1 &   3.88-4.24&& 8.2 &   3.73-4.36&& 8.3 &   3.50-4.50\\
  \hline 
  \end{tabular}%
  \end{center}%
  \footnote{4}{$1\le Z\le 92$ + {\sl ad--hoc} component, protons only.}\\
  \footnote{5}{$1\le Z\le 92$ only.}
\end{table*}

\end{appendix}

\end{document}